\documentclass[aps, prx, twocolumn, superscriptaddress, floatfix]{revtex4-2}\usepackage{mathrsfs} 
\usepackage{amsmath}
\usepackage[version=4]{mhchem}
\usepackage{amsfonts} 
\usepackage{amsmath,amssymb}
\usepackage{graphicx}
\usepackage{subfig}
\usepackage{float}
\usepackage[table]{xcolor}
\usepackage{colortbl}
\usepackage{tikz}
\usepackage{pgfplots}
\usepackage{bm}
\usepackage{multirow}
\usepackage[explicit]{titlesec}
\usepackage[most]{tcolorbox}
\usepackage{braket}
\usepackage{soul}
\usepackage{physics}
\usepackage{array}
\usepackage{booktabs}
\usepackage{hyperref}
\hypersetup{
    colorlinks=true,
    linkcolor=blue,
    filecolor=magenta,      
    urlcolor=red,
    citecolor=cyan,
    }
\pgfplotsset{compat=1.18}

\DeclareCaptionJustification{justified}{\leftskip=0pt \rightskip=0pt \parfillskip=0pt plus 1fil}

\begin{document}

\title{Kondo overscreening in the presence of superconductivity.}


\author{Pradip Kattel}
\email{pradip.kattel@rutgers.edu}

\author{Abay Zhakenov}

\author{Natan Andrei}
\affiliation{Department of Physics and Astronomy, Center for Materials Theory, Rutgers University,
Piscataway, NJ 08854, United States of America}

\begin{abstract}

We study the interplay and competition between Kondo overscreening and superconductivity. The model we consider consists of a single spin$-\frac{1}{2}$ quantum impurity at the edge of a quantum wire coupled to spin$-1$ bulk fermions that interact attractively, generating a (superconducting) mass gap.  The competition leads to a rich phase structure. We find that for strong Kondo coupling, there is a regime of phase space where the Kondo phase is stable with the impurity  \textit{overscreened} by a multiparticle Kondo effect, with a  Kondo scale dynamically generated. When the bulk and boundary interaction strengths are comparable, we find that a midgap state appears in the spectrum and screens the impurity, while in the ground state, the impurity is unscreened. This midgap state is akin to the Yu-Shiba-Rusinov (YSR) states that exist in the entire phase space in the BCS superconductor. When the bulk superconducting interaction strength is stronger than the boundary Kondo interaction strength, the impurity can no longer be screened. Further, between the Kondo and YSR phases, we find a novel phase where, while the Kondo cloud overscreens the impurity, a boundary excitation exists that has vanishing energy in the thermodynamic limit. Similar phase diagrams that  result from  competition between different mechanisms
were found for other models too: the dissipative Kondo system where dissipation competes with screening \cite{kattel2024dissipation}, the Kondo impurity coupled to spin-1/2 attractively interacting fermions \cite{pasnoori2022rise} where condensation competes with screening,  and the  XXX-Kondo model where the lattice cutoff and the bulk spin interaction compete with screening \cite{wang1997exact, kattel2023kondo,kattel2024spin}.

\end{abstract}
\maketitle

\section{Introduction}
The conventional Kondo effect~\cite{kondo2012physics,hewson1997kondo,yosida1996theory} is characterized by the screening, at low temperature, of a magnetic moment antiferromagnetically coupled to a non-interacting electron bath. It has been studied by several non-perturbative methods like Wilson's numerical RG~\cite{wilson1975renormalization,krishna1980renormalizationa,krishna1980renormalizationb}, the Bethe Ansatz~\cite{andrei1980diagonalization,vigman1980exact,andrei1983solution, tsvelick1983exact}, boundary conformal field theory~\cite{affleck1991kondo,affleck1995conformal,saleur2002lectures}, and more recently a combination of large S and renormalization group approach~\cite{krishnan2024kondo}. In the language of conformal field theory, the Kondo effect is captured by the  RG flow of the Kondo boundary defect coupling from the UV theory where a decoupled spin$-\frac{1}{2}$ impurity has a zero temperature entropy $S_{UV}=\ln 2$ to the IR theory where the impurity is completely screened by the conduction electrons and hence has entropy $S_{IR}=0$.

However, exact non-perturbative solutions are still scarce when the effect of a single quantum impurity is considered in strongly interacting quantum field theories describing many condensed matter systems of interest like superconductors~\cite{KScuevas2001kondo,KSmuller1971kondo,KSborkowski1992kondo,KSsteglich2016foundations,Shiba,Yu,Rusinov,pasnoori2022rise,pasnoori2020kondo} and spin liquids~\cite{SPINLIQUIDkim2008kondo,SPINLIQUIDchen2022exotic,florens2006kondo} or recently introduced particle physics problem of Kondo effect in quantum chromodynamics (QCD)~\cite{QCDPhysRevC.88.015201,QCDPhysRevD.92.065003} where a heavier quark (c,b) acts as an impurity in a quantum matter made up of light quarks (u,d,s) at extreme densities.

We begin by studying a single impurity in a strongly correlated 1+1-dimensional field theory described by an attractive   4-fermion  $O(3)$ invariant color interaction first studied in~\cite{andrei1984dynamical}. The O(3) model generalizes the standard SU(2)  Gross-Neveu model that was introduced as a model of elementary particles that is asymptotically free and where dynamical symmetry breaking takes place, giving mass to the fermions~\cite{gross1974dynamical,NJPhysRev.122.345,NJPhysRev.124.246,andrei1979diagonalization}.  Likewise, the model can be viewed as a model of superconductivity~\cite{liu2019superconductivity,thies2006relativistic,kalinkin2003phase,PRPPhysRevB.105.174517,marino1986relativistic} where the attractive interaction among left and right-moving fermions dynamically generates a superconducting mass gap $\Delta=2m$, where $m$ is the mass of a single spinon (kink) excitation.  However, since in $1+1$ dimensions the spontaneous breaking of a continuous symmetry \cite{MW-PhysRevLett.17.1133} and the existence of a Goldstone boson is not possible, according to Coleman's theorem \cite{coleman1973there}, the dynamical mass generation in this model is associated with a decoupled O(1) bosonic zero mode, which leads to power decay of the fermionic correlation functions \cite{witten1978chiral}.


The model considered in this work can also be viewed as a toy model of QCD Kondo or as a model of the Kondo effect in a one-dimensional superconducting wire. Using Bethe Ansatz, we shall provide exact non-perturbative results examining the role of bulk strong interaction on the Kondo effect. We address the $SU(3)$ Gross–Neveu model with impurity as a realistic description of the QCD Kondo problem in Ref.~\cite{Kattel:2025hgk}.

The Kondo effect in superconductors has been widely studied~\cite{Yu,Shiba,Rusinov,sakurai1970comments,franke2011competition,PhysRevLett.121.207701,aoi1974magnetic,PhysRevLett.26.428,matsuura1977effects,takano1969kondo,cuevas2001kondo}. In a superconductor, electrons interact attractively to form Cooper pairs throughout the bulk, establishing a coherent superconducting state. However, when an impurity is present, the conduction electrons also interact with the impurity to form a many-body Kondo singlet, and further, when the interaction with the impurity is strong enough, a local bound state may be formed that screens the impurity. The competition between the formation of Cooper pairs and the screening of the impurity by a singlet state leads to unexpected phenomena, including unconventional superconductivity and quantum critical behavior. A recent work studied a single impurity at the edge of an interacting $SU(2)$ superconducting wire and identified several impurity phases: the Kondo phase, where the impurity is screened by a multiparticle Kondo cloud; the Yu-Shiba-Rusinov (YSR) phase, where it is screened by a single-particle bound state; and an unscreened phase \cite{pasnoori2022rise}.

Here, we generalize this problem by considering a $SO(3)$ invariant superconducting bulk interacting with spin-$\frac{1}{2}$ impurity in the boundary, and we show that the impurity is \textit{overscreened}. We shall show, further, that for antiferromagnetic coupling, the impurity exhibits four different phases: i. the Kondo phase where it is \textit{overscreened} by a many-body Kondo cloud where the zero temperature entropy of the impurity is $\frac{1}{2}\ln 2$, ii.  the zero mode phase where the impurity is many-body overscreened and there also exists a zero energy boundary excitation so that the ground state is degenerate with the ground state in the Kondo phase in the thermodynamic limit, iii. the YSR phase, where the impurity is unscreened in the ground state but can be screened by a single particle bound mode formed at the edge of the wire, which has a finite mass below the bulk mass gap, and finally iv.  a local moment phase, where the impurity cannot be screened at any energy scale. Apart from the dynamically generated mass gap in the bulk $m$, the model also dynamically generates a Kondo scale $T_K$ in the Kondo and zero mode phases that govern the overscreening of the impurity.

\bigskip

\section{The Model}
The model Hamiltonian we consider is 

\begin{widetext}
    \begin{align}
    H&=\int_{-L}^0\mathrm{d}x  \sum_{r,s,u,v=1}^3 \left[-i (\psi^\dagger_{r+}\partial_x \psi_{r+}-\psi^\dagger_{r-}\partial_x \psi_{r-})+2g\, \psi^\dagger_{r+}\psi^\dagger_{s-}(\sum_\alpha \tau^\alpha \otimes \tau^\alpha+I)_{rs}^{uv} \psi_{u-}\psi_{v+}+2J \,\psi^\dagger_{r,-}(0)\vec \tau_{rs}\psi_{s,+}(0)\cdot\vec \sigma\right],
    \label{modelham}
\end{align}
\end{widetext}
with open boundary conditions
\begin{equation}
    \psi_+(0)=-\psi_-(0) \quad \text{and}\quad \psi_+(-L)=-\psi_-(-L)
\end{equation}
at both edges.
The matrices $\tau^\alpha$ denote the three $O(3)$ generators explicitly written as
    \begin{equation}
     \tau^x=\left(
\begin{array}{ccc}
 0 & 0 & 0 \\
 0 & 0 & -i \\
 0 & i & 0 \\
\end{array}
\right)\,  \tau^y=\left(
\begin{array}{ccc}
 0 & 0 & i \\
 0 & 0 & 0 \\
 -i & 0 & 0 \\
\end{array}
\right) \,\tau^z=\left(
\begin{array}{ccc}
 0 & -i & 0 \\
 i & 0 & 0 \\
 0 & 0 & 0 \\
\end{array}
\right),
\label{rep2}
\end{equation} 
$I_{ac}^{bd}=\delta_{a,b}\delta_{c,d}$ is the identity matrix,  $\vec \sigma$ are the $SU(2)$ Pauli matrices acting on the spin space of the spin-$\frac12$  impurity localized at $x=0$, $\pm$ denotes the chirality and $r,s=1,2,3$ is the $O(3)$ color index. 
 The coupling  $g>0$  describes the bulk interactions of both the attractive spin and charge degrees of freedom between the fermions of opposite chirality and the coupling  $J>0$ describes the chirality-changing interaction between the bulk fermion and the localized impurity at the right edge of the chain.

 When the boundary coupling $J$ is zero, the bulk model reduces to a Bethe Ansatz-solvable $O(3)$  \cite{andrei1984dynamical} and was studied in depth using both large $N$ analysis and integrability techniques. It was shown that a dynamical symmetry breaking takes place, and the model opens a mass gap and contains fractionalized kink excitations. On the other hand, when $g=0$, the model reduces to the Bethe Ansatz solvable case of the higher spin Kondo model studied in~\cite{tsvelick1985exact} where the spin-$\frac12$ impurity is overscreened by spin-$1$ bulk and exhibits behavior similar to the two-channel Kondo found in \cite{andrei1984solution}.
 
 Here, we shall show that when both $J$ and $g$ are non-zero, the model can still be solved exactly via the Bethe Ansatz approach ~\cite{bethe1931theory,sutherland2005introduction,baxter2016exactly,gaudin2014bethe,slavnov2022algebraic,vsamaj2013introduction, eckle2019models,franchini2011notes}. We shall explicitly construct the eigenstate of the Hamiltonian Eq.\eqref{modelham} in Appendix \ref{Nparteig-sec} and then prove that the model is still integrable in the presence of the boundary perturbation in Appendix \ref{integrability}. We shall derive the Bethe Ansatz equations of the model using the functional Bethe Ansatz method \cite{sklyanin1990functional,wang2015off} and fusion hierarchy \cite{wang2015off,mezincescu1992fusion,BABUJIAN1983317,Cao_2015} in Appendix \ref{integrability} and alternatively by using the fusion technique of \cite{andrei1984dynamical} in Appendix \ref{altBAE}.  
 
 Note that due to isomorphism $\mathfrak{su}(2)\cong\mathfrak{so}(3)$, it is also possible to view the bulk Hamiltonian as $SU(2)$ spin-$1$ Gross-Neveu model (see Appendix \ref{su2level2}). Since the boundary interaction is with spin-$\frac{1}{2}$ spin, it is often useful to use $SU(2)$ representation to label the states in the theory. 

 \bigskip

\begin{figure}
    \centering
    \includegraphics[width=1\linewidth]{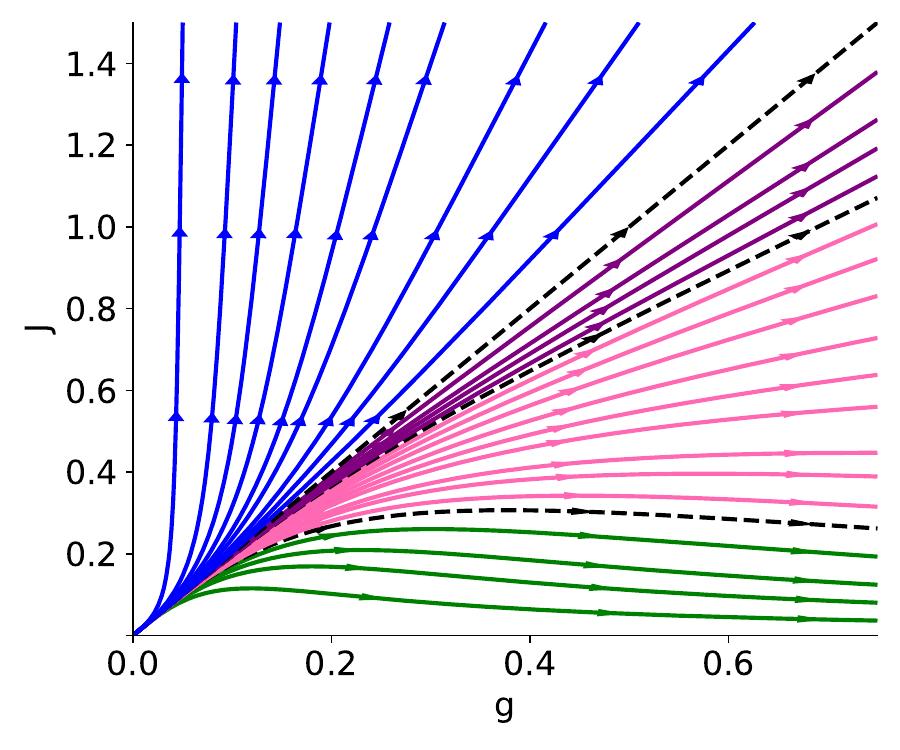}
    \caption{The weak-coupling RG flow diagram for Hamiltonian Eq.\eqref{modelham} is given by  Eq.\eqref{ggflow} and Eq.\eqref{jjflow}. The flow is away from the non-interacting point $J=0=g$. The blue curves denote the overscreened Kondo phase, where the boundary coupling flows to a non-trivial 2-channel Kondo fixed point. The purple lines represent the zero-mode phase, where the boundary coupling flows to the 2-channel fixed point, but there is a boundary excitation of zero energy.  Pink lines indicate the YSR phase, where the impurity is unscreened in the ground state but screened by single-particle bound modes in excited states. Green lines depict the unscreened regime, where the boundary coupling flows to zero, such that the impurity is unscreened. The dashed black curves demarcate the four boundary phases of the model.}
    \label{fig:RGplot-plot}
\end{figure}

In the language of the renormalization group (RG), the model Eq.~\eqref{modelham} presented here is characterized by two coupling constants, both of them flowing under the action of the RG.  Their weak coupling flow can be extracted perturbatively, but to determine their full structure, we need information provided by the exact solution.
 The RG flow diagrams are given by the RG equations
\begin{align}
   \beta(g)&=-\frac{2}{\pi}g^2 \label{ggflow}\\
    \beta(J)&=-\frac{2 J \left(2 g J^2-g+J\right)}{\pi  \left(2 J^2+1\right)}
    \label{jjflow}
\end{align}
 obtained from weak-coupling perturbation theory or 
 from the exact solution in Appendix \ref{RGderivation} \footnote{The exact RG equations obtained from the Bethe Ansatz solution are expressed in an unconventional cutoff scheme where the UV cut-off is imposed in the fully interacting theory. However, at weak coupling, both schemes coincide \cite{andrei1983solution}.}. Since $\beta(g)$ is always negative, $g$ always flows to the strong coupling where the  Hamiltonian Eq.\eqref{modelham} gains mass via dimensional transmutation. In contrast, $\beta(J)$ can change sign depending on the relative strengths of $J$ and $g$; therefore $J$ flows to strong coupling where the impurity is overscreened by many body Kondo cloud only when $J\gg g$,  but for $J\ll g$ it flows back to weak coupling, as shown in Fig.~\ref{fig:RGplot-plot}, and the impurity is not screened. In intermediate regimes where $g\sim J$, there exist two phases. One we call the zero mode phase, where the impurity is still overscreened by the many-body cloud, but a unique zero-energy boundary excitation is given by an isolated boundary string solution of the Bethe Ansatz equations. Further, there exists another phase that we call the YSR phase, which is similar to the Yu-Shiba-Rusinov phase in BCS superconductors, where the impurity is no longer screened in the ground state, but there is an elementary boundary excitation where a single particle in an edge-localized bound mode screens the impurity. 
 
 The phases
 correspond to definite ranges of the RG invariant $d(J,g)$
 \begin{equation}
    d= d(J,g)=\sqrt{\frac {1} {4 g^2} - \frac {1 - 2 J^2} {2 g J} - \frac {9} {4}}
\end{equation}
that remains constant while both $g$ and $J$ are running coupling constants that vary with the energy scale. This parameter could take either a real value or a purely imaginary value. In Fig.\ref{fig:RGplot-plot}, constant $d$ surfaces are shown via lines with arrows and denote the boundaries between the phases.  These phases are illustrated in Fig.\ref{fig:RGplot-plot}.  We shall also use this parameter $d$ to label various impurity phases in the model in the later sections. Notice that, at the diagonal line $J=2g$, the two equations (Eq.\eqref{ggflow} and Eq.\eqref{jjflow}) become the same, where the flow lines above represent the Kondo phase, whereas the lines below represent the other three phases.  The weak-coupling renormalization group equations (Eqs. \eqref{ggflow} and \eqref{jjflow}) (obtained exactly from the Bethe Ansatz analysis) describe how the couplings evolve from the non-interacting point ($g=J=0$). They reveal that, depending on the relative strength of $g$ and $J$, the boundary coupling $J$ can either grow strong or diminish to zero, indicating at least two different regimes. By diagonalizing the Hamiltonian and carefully analyzing the spectrum, we find that the full-phase diagram actually comprises four distinct boundary phases, each exhibiting a unique low-energy behavior. Although the RG equations reveal how the two couplings evolve, they do not determine whether the model flows to strong or weak coupling fixed points (if such fixed points exist). To pinpoint the infrared fixed points, we employ the Bethe Ansatz.

\section{Summary of results}

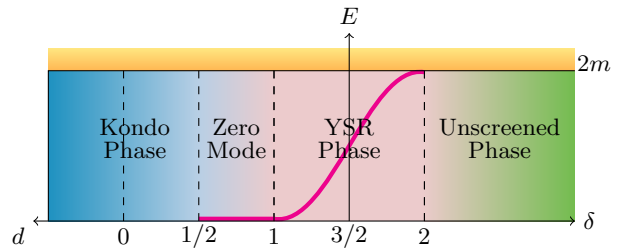
\begin{figure}
 \begin{tikzpicture}

\definecolor{phase1a}{RGB}{29,145,192} 
\definecolor{phase1b}{RGB}{184,207,229} 
\definecolor{phase2}{RGB}{236,203,204}  
\definecolor{phase3a}{RGB}{182,226,186} 
\definecolor{phase3b}{RGB}{115,188,79}  
\definecolor{topcolor1}{RGB}{255,230,128} 
\definecolor{topcolor2}{RGB}{255,182,82}  

\shade[left color=phase1a, right color=phase1b] (-1,0) rectangle (1,2);
\shade[left color=phase1b, right color=phase2] (1,0) rectangle (2,2);
\fill[phase2] (2,0) rectangle (4,2); 
\shade[left color=phase2, right color=phase3b] (4,0) rectangle (6,2); 

\shade[top color=topcolor1, bottom color=topcolor2] (-1,2) rectangle (6,2.3);

\draw[dashed] (2,0) -- (2,2);
\draw[dashed] (4,0) -- (4,2);

\draw[magenta,ultra thick] (1,0.03)--(2,0.03);
\draw[magenta,ultra thick] (2,0.03).. controls (2.75,-0.1) and (3.25,2.1).. (4,1.98);
\draw[dashed] (4,0) -- (4,2);

\draw[dashed] (0,0) -- (0,2);
\draw[dashed] (2,0) -- (2,2);
\draw[dashed] (1,0) -- (1,2);

\draw[<-] (-1.2,0) -- (-1,0);
\node at (-1.4, -0.2) {$d$}; 

\draw[->] (-1,0) -- (6,0) node[right] {$\delta$};
\node at (6.2,2.1) {$~2m$};
\draw[->] (3,0) -- (3,2.5) node[above] {$E$};
\draw (-1,0) -- (-1,2) -- (6,2);

\node at (2, -0.2) {1};
\node at (0, -0.2) {0};
\node at (1, -0.2) {1/2};

\node at (3, -0.2) {3/2}; 
\node at (4, -0.2) {2}; 

\node at (0.15, 1.25) {Kondo}; 
\node at (0.15, 0.95) {Phase}; 
\node at (1.5, 1.25) {Zero};
\node at (1.5, 0.95) {Mode};
\node at (3, 1.25) {YSR};
\node at (3, 0.95) {Phase};
\node at (5, 1.25) {Unscreened}; 
\node at (5, 0.95) {Phase}; 

\end{tikzpicture}
    \caption{The phase diagram of the spin-$1$ superconductor with a  spin $S = \frac{1}{2}$ Kondo impurity  at the boundary. It shows the spectrum dependence on $d$, the RG invariant parameter that combines both the bulk and boundary running coupling constants. In the Kondo regime, $d \in \mathbb{R}$ or $0<\delta < 1/2$ ($\delta=id$),  where the Kondo scale $T_K$ is greater than $m$ (The dynamically generated mass of the spinon), the impurity is overscreened by a multiparticle Kondo cloud. For the parametric range $1/2<\delta<1$, while the multiparticle Kondo cloud still overscreens the impurity, there exists a unique zero-energy excitation (shown by the solid magenta lines) in the thermodynamic limit which is described by a unique purely imaginary solution of the Bethe Ansatz equation $\lambda=\pm i\left(\frac{1}{2}-\delta\right)$. In the range, $1 < \delta < {2}$, the YSR phase, the purely imaginary solution, acquires finite mass $m\cos(\pi \delta)$ \textit{i.e.} corresponding to mid-gap states with energy shown in the solid magenta curve. In this phase, the impurity is no longer screened by the multiparticle cloud in the ground state but by a single particle bound state exponentially localized at the edge of the system. 
    Finally, for $\delta > {2}$, the impurity remains unscreened, and the mid-gap states are absent. The boundary string solution has zero energy in this regime, and it is not possible to add this solution to the ground state without adding an additional even number of holes \cite{pasnoori2022rise}. The orange-shaded region denotes the continuum of excitation above the mass gap $\Delta=2m$. Notice that this result is in sharp contrast with the classical impurity in the BCS superconductor, where the YSR phase exists in the entire phase space. The phase diagram shares similarities with the spin-1/2 Gross-Neveu model with a single Kondo impurity studied in \cite{pasnoori2020kondo,pasnoori2022rise} where the Kondo, YSR, and the unscreened phases appear, and the bound state in the YSR regime could have negative energy in the ground state.}
    \label{fig:phasediag}
\end{figure}

The bulk term of the model Hamiltonian Eq.\eqref{modelham} 
possesses several interesting properties such as dimensional transmutation, asymptotic freedom, and fractionalized excitations ~\cite{andrei1979diagonalization}. It describes a one-dimensional superconductor where the superconducting mass gap $\Delta=2m$ is dynamically generated and the two-point function of the fermionic bilinear $\psi^\dagger_{a-}\psi_{a+}(x)$ acquires a non-vanishing vacuum expectation value that falls off algebraically, thereby showing quasi-long-range order \cite{andrei1984dynamical}. When a spin-$\frac{1}{2}$ impurity is attached to the edge of the superconductor, it is the interplay between the bulk attractive interaction (that leads to superconducting order) and the boundary Kondo effect (where cloud of bulk fermion overscreens the impurity) that leads to a very rich impurity behavior: Both the superconducting coupling $g$ and the Kondo coupling $J$ are running coupling constants. In the absence of the Kondo coupling $J$, the superconducting coupling $g$ flows to the strong coupling and opens a mass gap in the spin sector of the theory, while the charge decouples and remains massless. Likewise, in the absence of the superconducting coupling $J$, the Kondo coupling $g$ flows to an intermediate non-Fermi liquid fixed point where the impurity is overscreened by the bulk fermions such that there is residual impurity entropy of $\frac{1}{2}\ln 2$ signaling a fractionalized degree of freedom (Majorana) at zero temperature. When both bulk superconducting coupling $g$ and the boundary Kondo coupling $J$ are present, they both flow. In such a case, we identify an RG invariant parameter $d$ that combines both the bulk and boundary running coupling constants. In terms of the RG invariant parameter $d$, the boundary physics can be classified into four distinct phases as shown in the boundary phase diagram Fig.~\ref{fig:phasediag}. When the parameter $d\in \mathbb{R}$ or when $d=i\delta$ and $0<\delta <\frac{1}{2}$, corresponding to large impurity coupling $J$ compared to the bulk interaction strength $g$ the impurity is in an overscreened Kondo phase where a multiparticle cloud of massive spinons in bulk overscreens the spin-1/2 impurity. A new Kondo scale $T_K$ is dynamically generated in this phase, which, among others, characterizes the overscreening cloud.  Moreover, there is residual impurity entropy of $\frac{1}{2}\ln 2$ just like in the conventional two-channel Kondo problem.
 When $\frac{1}{2}<\delta<1$, an intermediate phase appears where the impurity is still dynamically overscreened by the multiparticle cloud, but there is a unique boundary excitation where a boundary localized edge mode and a one-string solution of Bethe equations form a zero energy singlet excitation. This phase blends features of the Kondo phase (impurity screening) and the YSR phase (presence of a boundary excitation). So far, when the RG invariant parameter $d$ is real or when $d=i\delta$ is between $0<\delta<1$, the impurity is screened in the ground state by a multiparticle Kondo cloud.  At $\delta=1$, the model undergoes a boundary quantum phase transition, where for $\delta>1$, the impurity is no longer screened in the ground state. However, when $1<\delta<2$, a mid-gap state forms that can screen the impurity, albeit not in the ground state. Notice that an analogous mid-gap bound mode would exist throughout the phase space in a BCS superconductor with classical impurity \cite{Yu,Shiba,Rusinov} and screen it in the ground state. Here, however, the strong quantum fluctuation restricts it to a small region in the phase space.  Finally, when $\delta>2$, the bulk interaction strength is larger than the boundary interaction strength, and neither the Kondo physics is stabilized nor the midgap state survives in the spectrum. Thus, the impurity is no longer screened on any scale, and the model is in the local moment phase. 

 \begin{widetext}
\begin{table}
    \centering
    \renewcommand{\arraystretch}{1.3} 
    \begin{tabular}{|l|l|p{8cm}|}
        \hline
        \textbf{Phases} & \textbf{Parameter range} & \textbf{Properties} \\ 
        \hline
        Overscreened Kondo & $d\in\mathbb{R}\cup \delta\in (0,1/2)$ & 
        Impurity is overscreened by the multiparticle cloud. \\
        & & No boundary excitations. \\
        & & Kondo screening wins over the superconducting order. \\ 
        \hline
        Zero mode & $\delta \in (1/2,1)$ & 
        Impurity is overscreened by the multiparticle cloud. \\
        & & There exists a zero-energy boundary excitation. \\ 
        \hline
        YSR & $\delta \in (1,2)$ & 
        Impurity is not screened in the ground state. \\
        & & Impurity is screened in the mid-gap state. \\
        & & Finite energy boundary excitations. \\ 
        \hline
        Local moment & $\delta>2$ & 
        Impurity is not screened at any energy scale. \\
        & & Superconducting order wins over the Kondo screening. \\ 
        \hline
    \end{tabular}
    \caption{Four impurity phases and their key distinguishing properties.}
    \label{tab:impurity_phases}
\end{table}
\end{widetext}

 In summary, when the bulk interaction $g$ is absent, the model exhibits two-channel Kondo behavior throughout the phase for antiferromagnetic Kondo coupling $J$, where the impurity is overscreened by the Kondo screening cloud. Likewise, when the Kondo spin exchange interaction $J$ is absent, the model exhibits superconducting behavior where the attractive spin density-spin density interaction in the bulk opens a spin gap. When both the couplings are present the relative strengths of the two couplings lead to a rich boundary phenomenon. When $J\gg g$, the bulk exhibits superconducting order while at the boundary, the impurity is screened by a multiparticle Kondo cloud, just like in the conventional two-channel Kondo model. On the opposite limit, when $g\gg J$, the bulk superconductivity persists, but at the boundary, the Kondo impurity is entirely unscreened as if the Kondo coupling were ferromagnetic, which can be understood from the fact that the sign of the beta function in the flow equation Eq.\eqref{jjflow} changes to positive. There exist two intermediate phases when $g\sim J$ where in both cases the bulk has superconducting order, but in one case, the impurity is overscreened by a multiparticle cloud and in addition there exists a zero-energy boundary excitation,n whereas in another case, the impurity is unscreened in the ground state but there exists a mid-gap state where the impurity is screened by this single-particle state. We briefly tabulate the main distinguishing features between the four phases in Table \ref{tab:impurity_phases}

\section{Bethe Ansatz Equations}

We now turn to present the exact solution of the Hamiltonian Eq.\eqref{modelham}.
Since the scattering matrices of the model satisfy Yang-Baxter equations as shown in Appendix \ref{integrability}, there is no particle production or annihilation during scattering processes in this theory. Thus, it is possible to study the model by means of the Bethe Ansatz in the presence of a volume cut-off $L$, the length of the line segment with a fixed $N$ number of particles.   We derive the Bethe Ansatz equations using the fusion hierarchy \cite{frappatcomplete,Cao_2015,wang2015off} and Baxter's T-Q relation \cite{baxter1972partition,wang2015off} in Appendix \ref{integrability} in the presence of the volume cutoff $L$ and the UV cutoff $K=2\pi D$ where $D=\frac{N}{L}$ is the particle density.

The Bethe Ansatz equations we obtain for the model are (see Appendices \ref{integrability} and \ref{altBAE})
\begin{equation}
e^{-2ik_j L}=\prod_{\alpha=1}^M \left(\frac{ b+\lambda_\alpha+i}{ b+\lambda_\alpha-i} \right) \left(\frac{ b-\lambda_\alpha+i}{ b-\lambda_\alpha-i} \right)
\label{ccbae-final}
\end{equation}
where $k_j$ are the pseudomomenta, $M$ is the number of spin-flips and $\lambda_\alpha$ are the spin rapidities which satisfy following equations 
\begin{align}
&\prod_{\upsilon=\pm}\left(\frac{\lambda_\alpha+\upsilon  d+\frac{i}{2}}{\lambda_\alpha+\upsilon  d-\frac{i}{2}} \right)\left(\frac{\lambda_\alpha+\upsilon  b+i}{\lambda_\alpha+\upsilon  b-i} \right)^N\nonumber\\
&=\prod_{\beta\neq\alpha} \frac{\lambda_\alpha-\lambda_\beta+i}{\lambda_\alpha-\lambda_\beta-i}\frac{\lambda_\alpha+\lambda_\beta+i}{\lambda_\alpha+\lambda_\beta-i},
    \label{o3su21/2imp}
\end{align}
where $b=\frac{1}{2g}$ and $d$ is the RG invariant quantity we discussed earlier, related to the two bare parameters $g$ and $J$ as
\begin{equation}
     d=\sqrt{b^2-\frac{2 b}{c}-\frac{9}{4}} \quad \text{where} \quad c=\frac{2 J}{1-2 J^2}.
\end{equation}

Having solved for the momenta $k_j$, the energy of an eigenstate is then given by

\begin{equation}
E = \sum_j k_j.
\end{equation}

The problem of diagonalizing the Hamiltonian Eq.\eqref{modelham} thus requires consistently solving the coupled algebraic equations Eq.\eqref{ccbae-final} and Eq.\eqref{o3su21/2imp}.   The first equation,
 Eq.\eqref{ccbae-final} 
is a quantization condition for the pseudomomenta \(k_j\) (with \(L\) being the system size).  The second equation
\eqref{o3su21/2imp} constrains the spin rapidities \(\lambda_\alpha\). Every configuration of \(\lambda_\alpha\) that satisfies this equation corresponds to an eigenstate of the system, with its energy determined via the first equation. The spin of the state is obtained by knowing the total number of spin flips $M$ using the relation $ S^z=\left(N+\frac{1}{2}-2M\right)=\frac{1}{2}
$.

In essence, the first equation governs the charge degrees of freedom by fixing the allowed values of \(k_j\), while the second equation encodes the interactions among the spin degrees of freedom. Together, they capture the full spectrum of the model, with the interplay between charge and spin excitations determining both the eigenstates and their corresponding eigenenergies.

The first term in  Eq.\eqref{o3su21/2imp}, which depends on the RG invariant parameter $d$, encodes all the information about the boundary physics.  As we shall see, for various values of the parameter $d$, this term gives different contributions to a physical quantity, and hence, this gives us a way to describe the physics of the model in terms of the RG invariant parameter $d$, which could take real or imaginary values. We shall describe the boundary physics in four distinct phases: the Kondo phase, the zero mode phase, the YSR phase, and the unscreened phase with decreasing amount of details.

\

\section{The Overscreened Kondo phase}
When the RG invariant  $d$ is real or imaginary $d=i\delta$ with $0<\delta<1$, the impurity is overscreened by the Kondo cloud. The ground state of the model consists of a sea of two-string solutions $\lambda_\alpha=\chi_\alpha\pm \frac{i}{2}$ \cite{andrei1984dynamical} and a hole which corresponds to a spin-1/2 propagating spinon
in the ground state. The spinon is required for consistency reasons and counting of the states but does not participate in the impurity dynamics, as we shall briefly discuss later.

The Bethe equations for the centers of the two strings $\chi_\alpha$ take the form
\begin{widetext}
    \begin{align}
   &\prod_{\nu=\pm }\left(\frac{\chi_\alpha+\nu d+i}{\chi_\alpha+\nu d-i} \right) \left(\frac{\chi _{\alpha }+\nu b+\frac{3 i}{2}}{\chi _{\alpha }+\nu b-\frac{3 i}{2}}  \frac{\chi _{\alpha }+\nu b+\frac{i}{2}}{\chi _{\alpha }+\nu b-\frac{i}{2}}\right)^N \frac{\chi _{\alpha }+i}{\chi _{\alpha }-i}=\prod_{\nu=\pm}\prod_{\beta}^M\frac{\left(\chi _{\alpha }+\nu \chi _{\beta }+i\right)^2}{\left(\chi _{\alpha }+ \nu \chi _{\beta }-i\right)^2} \frac{\chi _{\alpha }+\nu \chi _{\beta }+2 i}{\chi _{\alpha }+ \nu \chi _{\beta }-2 i} . 
    \label{CBAE}
\end{align}

Taking the logarithm of both sides of the equation, we get
    \begin{align}
   \sum_{\nu=\pm}& N\arctan(2(\chi_\alpha+\nu b))+N\arctan(2/3(\chi_\alpha+\nu b)) +  \arctan(\chi_\alpha+\nu d)+\arctan(\chi_\alpha)-\pi I_\alpha \nonumber\\
   &=\sum_{\nu=\pm}\sum_\beta 2 \arctan(\chi_\alpha+\nu \chi_\beta)+\arctan(\frac12(\chi_\alpha+\nu \chi_\beta)),
    \label{FBAE}
\end{align}

where $I_\alpha$ are integers, which are the spin quantum numbers of the state.

Likewise, taking logarithms on both sides of Eq.\eqref{ccbae-final}, we obtain
\begin{equation}
E=\sum_j k_j=\sum_j \frac{\pi n_j}{L}+D  \sum_{\alpha=1}^M\left(\arctan( b+\lambda_\alpha)+\arctan( b-\lambda_\alpha) \right),
\label{maineng}
\end{equation}
which, for 2-strings, can be written by
substituting $\lambda_\alpha=\chi_\alpha\pm \frac{i}{2}$ as
        \begin{align}
    k_j=\frac{\pi n_j}{L}+\frac{1}{L}\sum_{\nu=\pm}\sum_{\alpha=1}^M \left[\tan^{-1}\left(2(b+\nu \chi_\alpha) \right)+\tan^{-1}\left(\frac{2}{3}(b+\nu \chi_\alpha) \right)\right],
    \label{engeng}
\end{align}
where $n_j$ are the charge quantum numbers. 

To analyze Eq.\eqref{FBAE} in the thermodynamic limit, we define the density of Bethe roots as
\begin{equation}
\rho(\chi_\alpha)=\frac{1}{\chi_{\alpha+1}-\chi_\alpha},
\end{equation}
    such that we convert the sums over $\alpha$ in Eq.\eqref{engeng} and Eq.\eqref{FBAE} into integral over $\chi$ as

\begin{align}
2\rho_0(\chi)&=f(\chi)-\int_{-\infty}^\infty\mathrm{d}\chi'~ K(\chi-\chi')\rho_0(\chi')-\int_{-\infty}^\infty\mathrm{d}\chi'~ K(\chi+\chi')\rho_0(\chi')+\mathcal{O}\left(\frac{1}{N}\right)
\label{SolDensity}
\end{align}
Since we are interested in the thermodynamic limit $N\to \infty$, the higher order terms are negligible. Here,
\begin{align}
f(\chi)&=\sum_{\nu=\pm}Na_{\frac12}(\chi+\nu b)+N a_{\frac32}(\chi+\nu b) + a_1(\chi+\nu d)+a_1(\chi),\nonumber\\
a_\gamma(\chi)&=\frac{1}{\pi}\frac{\gamma}{\chi^2+\gamma^2},\\
K(\chi)&=\frac{2}{\pi  \left(\chi ^2+1\right)}+\frac{2}{\pi  \left(\chi ^2+4\right)}=2a_1(\chi)+a_2(\chi)\label{kernel}.
\end{align}

Likewise, the energy can be computed from the root density using
\begin{equation}
    E=\sum_j \frac{\pi n_j}{L}+ D\int \rho_0(\chi)\left[\tan^{-1}\left(2(b+\nu \chi_\alpha) \right)+\tan^{-1}\left(\frac{2}{3}(b+\nu \chi_\alpha) \right)\right]\mathrm{d}\chi.
\end{equation}

Removing the $\chi=0$ solution, which is a trivial solution that results in vanishing wavefunction and adding a hole at position $\chi=\theta$, the solution of Eq.\eqref{SolDensity} is immediate in the Fourier space
\begin{align}
\tilde\rho_{gs}(\omega)&=\frac{2N \cos(b\omega)(e^{-\frac{| \omega | }{2}}+e^{-\frac{3| \omega | }{2}})+(2\cos(d\omega)+1)e^{-|\omega|}}{2 \left(1+e^{-| \omega | }\right)^2}-\frac{1+2\cosh(\omega\theta)}{2 \left(1+e^{-| \omega | }\right)^2}.
\label{rootdensF}
\end{align}
\end{widetext}

The total number of roots is then given by
\begin{equation}
    2 \tilde\rho_{gs}(0)=2\left(\frac{4N+3-1-2}{8}\right)=\frac{4N}{4}=N\in \mathbb{Z}
    \label{rootnumber}
\end{equation}
and the total spin of this state is
\begin{equation}
    S^z=\left(N+\frac{1}{2}-2M\right)=\frac{1}{2}
\end{equation}
Due to $SU(2)$ symmetry of the model, there exists another degenerate state with spin $S^z=-\frac{1}{2}$. Notice that in the absence of the spin-$\frac{1}{2}$ impurity at the edge, the ground state is a $S^z=0$ state~\cite{andrei1984dynamical}. The double-degenerate ground state with $S^z = \pm \frac{1}{2}$ arises from the presence of a hole in the ground state, which carries spin-$\frac{1}{2}$, while the impurity is overscreened by a multiparticle cloud. Notice that the impurity contribution to the density of root in Fourier space in the ground state is $\rho_{\mathrm{imp}}(\omega)=\frac {e^{-| \omega |}\cosh (d\omega)} {\left (e^{-| \omega |} + 
      1 \right)^2}$. Thus, the integrated density of state contribution due to impurity is $\int \rho_{\mathrm{imp}}(\lambda)\mathrm{d}\lambda=\frac{1}{4}$. The positive contribution shows that the impurity is screened. We shall explicitly demonstrate through an analysis of thermodynamics that the impurity is overscreened in Section \ref{therm}. It is important to recall that the last term in the density of hole Eq.\eqref{rootdensF}, which depends on $\theta$, is due to the presence of a hole in the ground state. This hole is required because the total number of roots given by Eq.\eqref{rootnumber} is not an integer.  We shall set the free parameter $\theta$ to 0 to minimize the energy to obtain the ground state. Notice that a similar hole would be present in the conventional Kondo problem ~\cite{andrei1983solution} if it is solved with an even number of electrons and there are no holes in the ground state for an odd number of electrons. However, in the current model, the hole exists in the ground state for both even and odd parity of a total number of fermions in the bulk due to the fact that the bulk is constructed by two copies of spin-$\frac{1}{2}$ models with equal number of bulk particles (see \cite{andrei1984dynamical} and Appendix \ref{altBAE}) and hence with a spin-$\frac{1}{2}$ impurity at the edge, there are effective odd numbers of spin-$\frac12$  of particles in the model.

\subsection{Excitations}\label{excitations}
In the Kondo regime (i.e., when $d\in R$ or $d=i\delta$ such that $0<\delta<\frac12$ shown in the phase diagram Fig.\ref{fig:phasediag}), there are no boundary excitations; all the excitations are bulk excitations. A thorough discussion of the bulk excitation can be found at \cite{andrei1984dynamical}; here, we shall briefly discuss it.  The simplest excitations involve spinless charge excitations. 

\textbf{Charge excitations (particle-hole)} – obtained by exciting the charge degrees of freedom that do not change spin quantum numbers. Starting from given \( n_j^0 \), where \(-K \leq \left(\frac{2\pi}{L}\right) n_j^0 < 0\), if quantum number is changed to \( n_j' = n_j^0 + \Delta n \geq 0\), the change in energy involved is
\begin{equation}
    \Delta E_c = \frac{2\pi}{L} \Delta n > 0.
\end{equation}
Here, $K>\left|\frac{2\pi}{L}n_j\right|$ is a cut-off imposed on the fully interacting theory to define the `bottom of the sea' with respect to which we study the excitations. 

\textbf{Spin exciations} - are constructed by changing the spin quantum numbers $I_\gamma$. For example, elementary spin excitation can be constructed by adding a pair of holes in the sea of two strings in position $\theta_1$ and $\theta_2$ and a one-string solution at position $\lambda$. The change in the root density is given by
\begin{align}
   \Delta \tilde\rho_T(\omega)=-\frac{\sum_{\eta=1}^2\cosh(\omega \theta_\eta)+ \left(e^{-\frac{3}{2} (| \omega | )}+e^{-\frac{| \omega | }{2}}\right) \cos (\lambda  \omega )}{ \left(1+e^{-| \omega | }\right)^2}
\end{align}

The spin of this excitation is 
\begin{equation}
    S^z_T=1,
\end{equation}
and the energy of this excitation is given
\begin{equation}
    E_T=D\sum_{\upsilon=\{1,2\}} \tan^{-1}\left( \frac{\cosh(\pi \theta_\upsilon)}{\sinh(\pi b)}\right)
\end{equation}
 The minimum energy of a single spinon occurs at $\theta=0$
\begin{equation}
    m=D\arctan\left(\frac{1}{\sinh(\pi b)} \right)
\end{equation}
In the scaling limit $\{D,b\}\to \infty$ while holding $m$ fixed, the energy of the triplets can be written as
\begin{equation}
    E_T=m\cosh(\pi \theta_1)+m \cosh(\pi\theta_2).\label{etrip}
\end{equation}

It is important to note that the massless bare particles have now acquired physical mass (m), and the Lorentz invariance of the model is restored in the scaling limit as evident from the energy expression in terms of the rapidity $\theta_i$ in Eq.\eqref{etrip}. 

We shall skip the details presented in~\cite{andrei1984dynamical} and describe other fundamental excitations. By creating two holes in the sea of two-strings and adding one one-string and one three-string solution, we obtain a singlet excitation with spin $S^z_S=0$ and energy that is the same as that of triplets in the thermodynamic limit \textit{i.e.}
\begin{equation}
    E_S=E_T=m\cosh(\pi \theta_1)+m \cosh(\pi\theta_2).
    \label{esing}
\end{equation}

Two more fundamental excitations can be constructed in the different fermionic parity sectors by changing the number of charge excitations by 1 and thereby creating a massless particle of energy $|q|$. The triplet excitation is constructed by adding two holes in the sea of two strings, and the singlet is constructed by adding one three-string solution on top of the two-string sea and two holes. In the thermodynamic limit, both of these excitations have the same energy
\begin{equation}
    E=|q|+m\cosh(\pi \theta_1)+m \cosh(\pi\theta_2).
\end{equation}

All other excitations can be constructed by adding more bulk string solutions, even the number of holes, quartet solutions, etc.~\cite {destri1982analysis}.

\subsection{Density of states}
From the root density in in Eq.\eqref{rootdensF}, we compute the density of states using the relation $\rho_{\mathrm{DOS}}(E)=2\left|\rho_{\mathrm{gs}}(\chi)/E'(\chi)\right|$ where $\rho_{gs}(\chi)$ is the inverse Fourier transform root density in the ground state given in Eq.\eqref{rootdensF}. Notice that the root density naturally separates into $\rho_{\mathrm{gs}}(\chi)=N\rho_{\mathrm{bulk}}(\chi)+\rho_{\mathrm{imp}}(\chi,d)+\rho_{\mathrm{boundary}}(\chi)$.

The density of state contribution from the bulk is
\begin{equation}
    \rho_{\mathrm{DOS}}^{\mathrm{bulk}}(E)=\frac{E/\pi}{\sqrt{E^2-m^2}},
\end{equation}
\begin{widetext}
   typical of a superconductor, and the impurity contribution becomes
\begin{equation}
    \rho_{\mathrm{DOS}}^{\mathrm{imp}}(E)=\frac{m \cosh (\pi  d) \left(E^2-m^2\right) \cosh ^{-1}\left(\frac{E}{m}\right)-\pi  d E m \sinh (\pi  d) \sqrt{E^2-m^2}}{2 \pi ^2 \left(m^2-E^2\right) \left(m^2 \cosh ^2(\pi  d)-E^2\right)},
    \label{kondoimpdos}
\end{equation}
such that the ratio of the impurity to the bulk per unit length contribution becomes
\begin{align}
    R(E)&=\frac{m \cosh (\pi  d) \sqrt{E^2-m^2} \cosh ^{-1}\left(\frac{E}{m}\right)-\pi  d E m \sinh (\pi  d)}{\pi  E \left(E^2-m^2 \cosh ^2(\pi  d)\right)},\quad \text{where} \quad E>m.
    \label{redef}
\end{align} 

\end{widetext}

\begin{figure}[H]
    \centering
    \includegraphics[width=\linewidth]{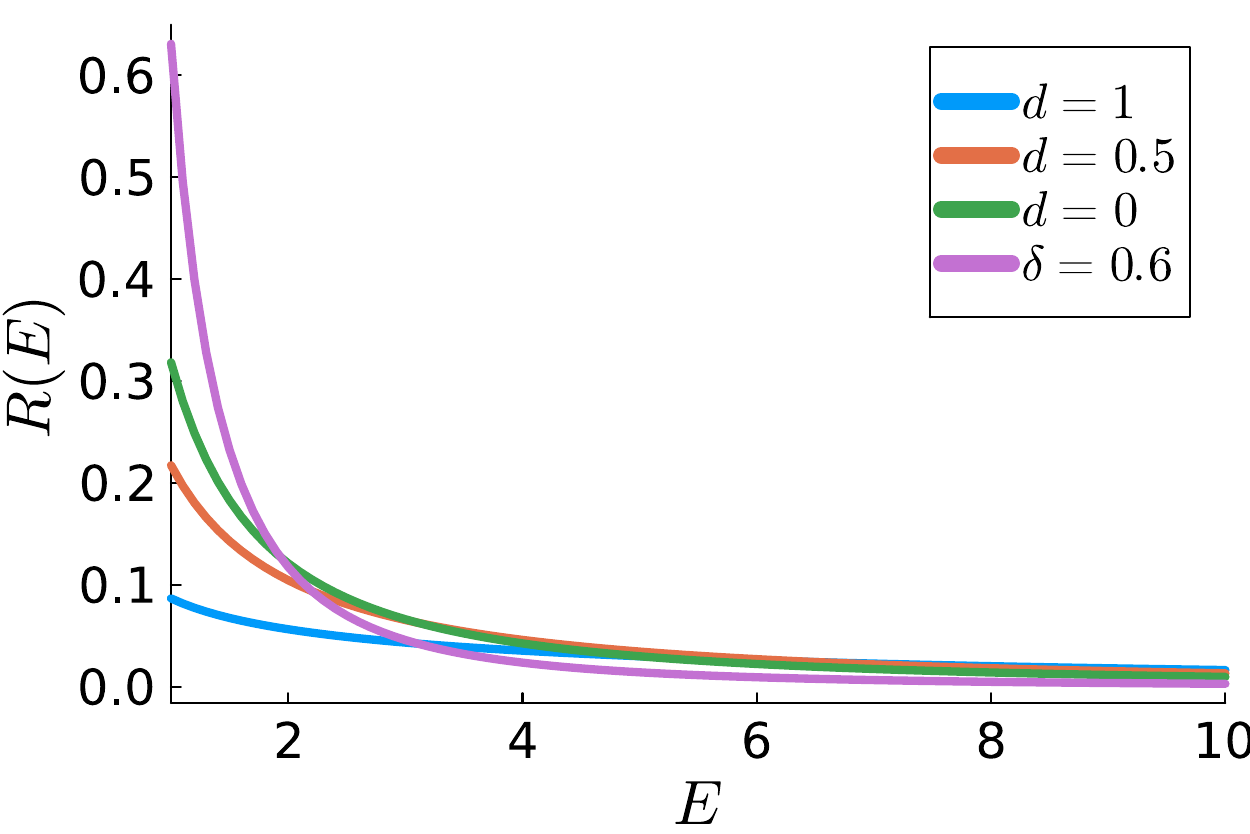}
    \caption{$R(E)$ for $m=1$ and various values of $d$ and $\delta=id$ shows the familiar peak in the ratio of the density of states in the ground states of Kondo and Zero mode phases.}
    \label{fig:RE}
\end{figure}
The impurity Kondo phase is characterized by a dynamically generated, RG invariant energy scale $T_K$ characterizing its response to external probes such as temperatures or magnetic fields. It is obtained via
\begin{equation}
    \int_m^{T_K} d E \rho^{\mathrm{imp}}_{\mathrm{DOS}}(E)=\frac{1}{2} \int_m^{\infty} d E \rho^{\mathrm{imp}}_{\mathrm{DOS}}(E),
    \label{Tkdef}
\end{equation}

which gives
\begin{equation}
    T_K=mf(d)
\end{equation}
where $f(d)$ is a function of the RG invariant parameter, which we obtain by graphically solving Eq.\eqref{Tkdef}
\begin{figure}
    \centering
    \includegraphics[width=0.5\textwidth]{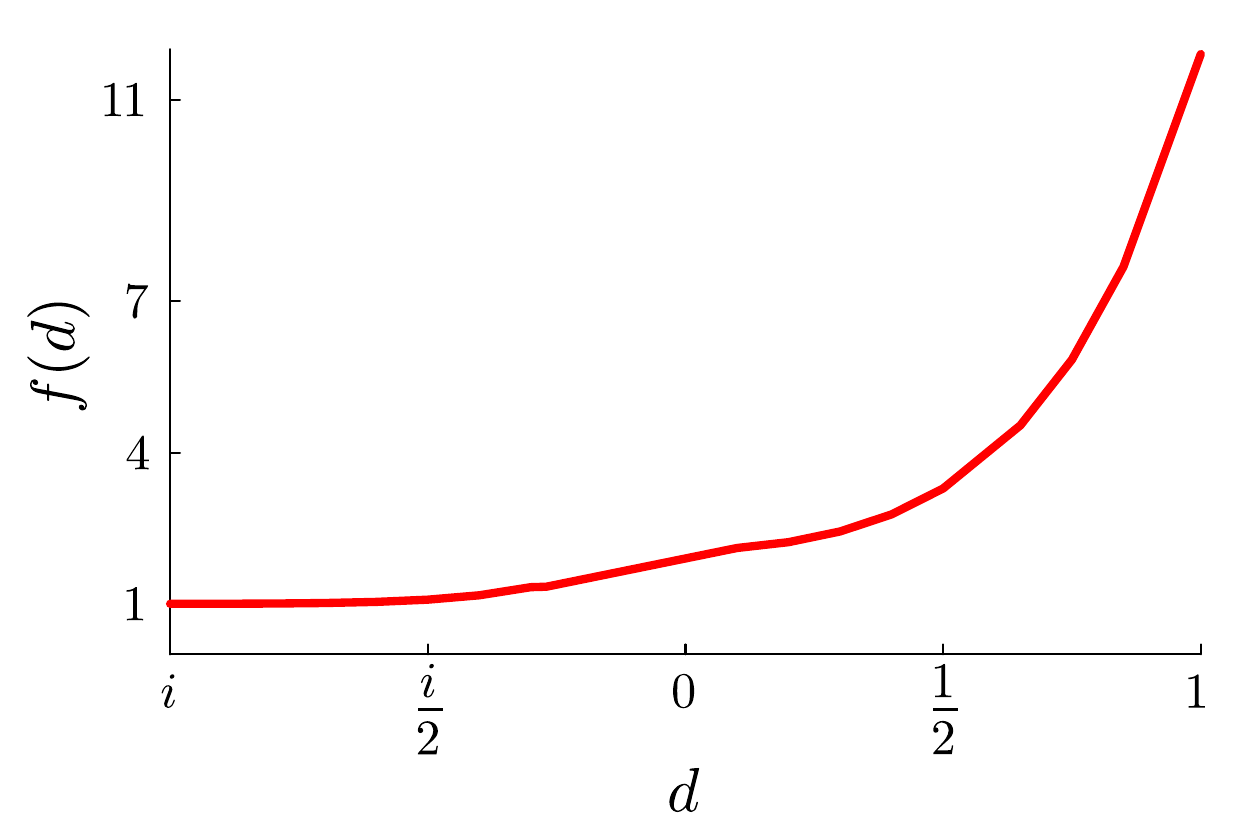}
    \caption{The plot of $f(d)=T_K/m$ as a function of the RG invariant parameter $d$ in the Kondo and the zero mode phase. Notice that $T_K\geq m$ for all values of $d$, and the minimum $T_K=m$ occurs at $d=i\delta=i$, which is the boundary between the zero mode phase and the YSR phase.}
    \label{fig:ffunc}
\end{figure}

 In summary, \textit{the overscreened Kondo phase} is characterized by the many-body overscreening of the spin-\(\frac{1}{2}\) impurity by the strongly interacting bath of spin-1 electrons. This leads to an emergent non-Fermi liquid fixed point, similar to the two-channel Kondo effect, where the impurity entropy at low temperatures takes the value \(\frac{1}{2} \ln 2\) as we shall see below. A dynamically generated Kondo scale \(T_K\), distinct from the bulk superconducting mass gap \(\Delta\),  governs the screening process. In the bulk, the attractive interaction between fermions leads to the formation of a superconducting order with a dynamically generated mass gap \(\Delta = 2m\). This boundary phase is stable for sufficiently large Kondo coupling \(J\), where the boundary interaction dominates over the bulk superconducting correlations, allowing the impurity to be overscreened by a multiparticle Kondo cloud.

\section{The zero mode Phase}
When the impurity parameter $d$ takes the purely imaginary value $d= i\delta$ where $\frac{1}{2}<\delta<1$, the impurity is in the zero mode phase as shown in the phase diagram Fig.\ref{fig:phasediag}. In this regime,
 the ground state is composed of  2-strings and a hole, with the root density given by 
\begin{align}
\tilde\rho_{gs}(\omega)&=\frac{2N \cos(b\omega)(e^{-\frac{| \omega | }{2}}+e^{-\frac{3| \omega | }{2}})+(2\cosh(\delta\omega)+1)e^{-|\omega|}}{2 \left(1+e^{-| \omega | }\right)^2}\nonumber\\
&-\frac{1+2\cosh(\omega\theta)}{2 \left(1+e^{-| \omega | }\right)^2}.
\label{rootdensZM}
\end{align}
Just as in the Kondo phase, the impurity is overscreened by the multiparticle Kondo cloud, and the spin of the ground state is 
\begin{equation}
    S^z=\pm \frac{1}{2},
\end{equation}
which is due to there being a spin-$\frac{1}{2}$ propagating spinon in the ground state. The ground state is an analytic continuation of the Kondo phase and is characterized by the dynamically generated scale $T_K=mf(\delta)$. The bulk excitations are also the same as in the Kondo phase, and are constructed by adding even numbers of holes in bulk strings as described in Sec.\ref{excitations}. The impurity density of states in the ground state is the analytic continuation of the density of states in the Kondo phase given by Eq.\eqref{kondoimpdos} to imaginary $d=i\delta$.

However, there also exists a unique boundary excitation in this phase - a boundary-bound mode with zero energy. This mode is described by the purely imaginary solution of the Bethe Ansatz equations Eq.\eqref{o3su21/2imp}. More precisely, when the RG invariant parameter $d=i\delta$ is purely imaginary and $\delta>\frac{1}{2}$, the Bethe equation Eq.\eqref{o3su21/2imp} has a purely imaginary solution
\begin{equation}
    \lambda_\alpha=\pm  i\left(\frac{1}{2}-\delta\right).
    \label{boundary-string-sol}
\end{equation}
The purely imaginary Bethe roots, associated with bound states, appear as poles in the physical or dressed boundary scattering matrix \cite{BSghoshal1993bound,BSskorik1995boundary,BSghoshal1994boundary,corrigan10094boundary}. This kind of  boundary string solutions describing the boundary excitation exist in various integrable systems with boundaries\cite{kapustin1996surface,pasnoori2020kondo,kattel2023exact,kattel2024dissipation,pasnoori2022rise,kattel2023exact,rylands2020exact,wang1997exact,chen1998open,chen1999pinning}.

When $1>\delta>\frac{1}{2}$, the change in the root density due to this root is
\begin{equation}
    \Delta_d\rho(\omega)=-\frac{e^{\delta  | \omega | }+e^{-((\delta -2) | \omega | )}+2 e^{| \omega | } \cosh (\delta  \omega )}{2 \left(e^{| \omega | }+1\right)^2}
\end{equation}
such that the total energy contribution due to this solution obtained from Eq.\eqref{maineng}
is exactly equal but with the opposite sign to that of its bare energy $D\tan ^{-1}\left(\frac{b}{\frac{3}{2}-\delta }\right)+D\tan ^{-1}\left(\frac{b}{\delta +\frac{1}{2}}\right)$. Hence, the total energy of this excitation vanishes in the thermodynamic limit. 

The state  is degenerate with the ground state in the thermodynamic limit is constructed by adding a boundary string solution and a one-string at position $\lambda_c$, resulting in a root density of the following form
\begin{widetext}
    \begin{align}
\tilde\rho_{gs}(\omega)&=\frac{2N \cos(b\omega)(e^{-\frac{| \omega | }{2}}+e^{-\frac{3| \omega | }{2}})+(2\cosh(\delta\omega)+1)e^{-|\omega|}}{2\left(1+e^{-| \omega |  }\right)^2}-\frac{1+2\cosh(\omega\theta)}{2 \left(1+e^{-| \omega |  }\right)^2}\nonumber\\
&-\frac{e^{\delta  | \omega | }+e^{-((\delta -2) | \omega | )}+2 e^{| \omega | } \cosh (\delta  \omega )}{2 \left(e^{| \omega | }+1\right)^2}-\frac{ \left(e^{-\frac{3}{2} (| \omega | )}+e^{-\frac{| \omega | }{2}}\right) \cos (\lambda_c  \omega )}{ \left(1+e^{-| \omega |  }\right)^2}
\label{rootdensZM-ex}
\end{align} 
\end{widetext}

The spin of this state is $S^z=\pm \frac{1}{2}$ (carried by the spinon), which is the same as that of the ground state because the one-string at position $\lambda_c$ and the boundary string $\lambda_\alpha$ form a singlet. Since both the boundary string and the one-string have vanishing energy in the thermodynamic limit, this two-fold state is degenerate with the ground state, as mentioned above. 

The position of the one string at $\lambda_c$ is fixed by the Bethe Ansatz equation written for this root, which takes the form
\begin{widetext}
    \begin{align}
&\left(\frac{\lambda_c+i/2}{\lambda_c-i/2} \right)\frac{\left(\lambda_c-i \left(\frac{3}{2}-\delta \right)\right) \left(\lambda_c-i \left(\delta-\frac{1}{2} \right)\right)}{\left(\lambda_c+i \left(\frac{3}{2}-\delta\right)\right) \left(\lambda_c+i \left(\delta-\frac{1}{2} \right)\right)} \left(\frac{\lambda_c-b+i}{\lambda_c-b-i} \right)^N\left(\frac{\lambda_c+b+i}{\lambda_c+b-i} \right)^N\nonumber\\
&=-\prod_{\beta} \frac{\lambda_c-\chi_\beta+3i/2}{\lambda_c-\chi_\beta-3i/2}\frac{\lambda_c-\chi_\beta+i/2}{\lambda_c-\chi_\beta-i/2}\frac{\lambda_c+\chi_\beta+3i/2}{\lambda_c+\chi_\beta-3i/2}\frac{\lambda_c+\chi_\beta+i/2}{\lambda_c+\chi_\beta-i/2}.
\label{post1st}
\end{align}
Using the root density Eq.\eqref{rootdensZM-ex} for the centers of the two strings $\chi$, we solve Eq.\eqref{post1st} in the thermodynamic limit and obtain that the position of the one-string $\lambda_c$, the position of the hole $\theta$, and the impurity parameter $\delta$ are related via
\begin{equation}
    \lambda_c= \frac{1}{\pi}\mathrm{arcsinh} \sqrt{\cos\pi \delta + \cosh \pi\theta +\cos \pi\delta \cosh \pi\theta}.
\end{equation}

Notice that even though the ground state and the zero-energy excited state are degenerate in the thermodynamic limit, these are physically quite distinct states. As mentioned earlier, in the ground state, the density of the state is the analytic continuation of that of the Kondo ground state. In the zero energy excited state, the density of states in the zero energy excited state contribution from the impurity, one string, and the boundary string is given by
\begin{align}
\rho_{\mathrm{DOS}} &= 
\frac{
    m \cos(\pi \delta) \left(E^2 - m^2\right) \cosh^{-1}\left(\frac{E}{m}\right) 
    + \pi \delta E m \sin(\pi \delta) \sqrt{E^2 - m^2}
}{
    2 \pi^2 \left(m^2 - E^2\right) \left(m^2 \cos^2(\pi \delta) - E^2\right)
} \nonumber \\
&-\frac{
    m \left[ 
        \pi (3\delta - 4) E \sin(\pi \delta) 
        + 3 \cos(\pi \delta) \sqrt{E^2 - m^2} \cosh^{-1}\left(\frac{E}{m}\right)
    \right]
}{
    \pi^2 \sqrt{E^2 - m^2} \left(-2 E^2 + m^2 \cos(2\pi \delta) + m^2\right)
}\nonumber\\
& -\frac{\text{sech}\left(\pi  \lambda_c -\cosh ^{-1}\left(\frac{E}{m}\right)\right)+\text{sech}\left(\pi  \lambda_c +\cosh ^{-1}\left(\frac{E}{m}\right)\right)}{4 \pi  \sqrt{E^2-m^2}}+ \delta(E).
\end{align}
Here, the first term is the analytic continuation of the impurity contribution in the Kondo phase, the second term is the contribution from the boundary string solution, the third term is a contribution from the one string, and the delta function is from the zero-energy states. Notice that the integrated density of state contribution from each the string solution and one string is $-\frac12$, which is exactly canceled by the delta function term, and the other first term has integrated density of state contribution of $\frac{1}{4}$, which shows that the impurity is screened in this state.
\end{widetext}

In summary, the \textit{zero mode phase} is an intermediate phase that exhibits characteristics of both the Kondo phase and the YSR phase. As in the former, the impurity is many-body screened by a multiparticle Kondo cloud, while in the latter, there exists a boundary excitation, albeit with vanishing energy in the thermodynamic limit. As the system parameters are tuned such that bulk superconducting strength competes with the Kondo screening mechanism, this phase emerges, retaining the overscreening mechanism of the Kondo phase but introducing an additional boundary excitation. The phase is characterized by the presence of a purely imaginary Bethe root solution, corresponding to a zero-energy localized mode at the boundary. The presence of this additional zero-energy excitation differentiates this phase from the conventional overscreened Kondo regime, marking a novel impurity behavior that arises due to the interplay of bulk and boundary interactions.

\section{The YSR Phase}
As mentioned earlier, when $\delta<1$, the impurity is screened in the ground state by a multiparticle cloud of massive spinons, but at $\delta=1$, the model undergoes a boundary quantum phase transition, and for $\delta>1$, the impurity is no longer screened in the ground state. In this section, we shall discuss 
 the regime $1<\delta<2$ dubbed as the YSR phase in the phase diagram Fig.\ref{fig:phasediag}. In this phase,   the boundary string solution described in Eq.\eqref{boundary-string-sol} acquires mass and gives rise to a mid-gap state above the ground state akin to the YSR state in the BCS superconductor, and in this mid-gap state, the impurity is screened \footnote{Notice that this is in contrast with the model with spin-$\frac{1}{2}$ bulk studied in \cite{pasnoori2022rise} where the YSR state is the ground state for some range of parameters -- here the YSR regime is a mid-gap excited state throughout the parametric regime $1<\delta<2$}. As we shall show below, this state requires an additional hole on top of this boundary string solution such that the energy of this state is below the mass gap. In the next section, we shall discuss the region $\delta>2$  where the impurity cannot be screened at any energy scale, though the boundary string solution exists and is gapless. But because the boundary string becomes a wide string \cite{destri1982analysis}, this solution cannot be added to the ground state without adding additional massive holes to form an excitation. Such an excitation would, therefore, be above the mass gap.

Here, for  $1<\delta<2$, the ground state is made up of 2-string solutions, which have the root density
\begin{widetext}
    \begin{equation}
    \tilde\rho_{\mathrm{2-string}}(\omega)=\frac{2N e^{-\frac{| \omega | }{2}}\cos(b\omega)+2N e^{-\frac{3| \omega | }{2}}\cos(b\omega)-2 e^{-\delta  | \omega | } \sinh (| \omega | )+e^{-|\omega|}-1}{2 \left(2e^{-| \omega | }+e^{-2| \omega | }+1\right)}.
    \label{rootdens2strings-nn}
\end{equation}
\end{widetext}
The impurity is unscreened in the ground state as $\int_{-\infty}^\infty \rho_{\mathrm{imp}}(\lambda)\mathrm{d}\lambda=0$, where $\rho_{\mathrm{imp}}(\lambda)$ is the root density contribution from the impurity (i.e., the $\delta$ dependent term). The spin in the ground state is 
\begin{equation}
    S^z=\left(N+\frac{1}{2}\right)-2\tilde\rho_{\mathrm{2-string}}(0)=\frac{1}{2}.
\end{equation}
Due to $SU(2)$ symmetry, there is another state degenerate to the ground state with spin $S^z=-\frac{1}{2}$.

In this phase, apart from the bulk excitations that are constructed by adding an even number of spinons, bulk strings of various lengths, etc, there is a unique boundary excitation where the impurity is screened by a single-particle bound mode. Consider an excitation consisting of a hole, along with the one-string solution $\lambda_c$ and the boundary string $\lambda_\alpha = \pm i\left(\frac{1}{2} - \delta\right)$, situated on top of the two-string solutions forming the ground state.
 Thus, the total root density of this state reads
\begin{widetext}
\begin{align}
    \tilde\rho_d(\omega)&=\frac{2N e^{-\frac{| \omega | }{2}}\cos(b\omega)+2N e^{-\frac{3| \omega | }{2}}\cos(b\omega)+e^{-|\omega|}-1-e^{(\delta -2) | \omega | }-e^{-\delta  | \omega | }-2\cos(\omega\theta)}{2 \left(1+e^{-| \omega | }\right)^2}
\nonumber\\
    &-\frac{2\cos(\omega\lambda_c)\left(e^{-\frac{3}{2} (| \omega | )}+e^{-\frac{| \omega | }{2}}\right) }{2 \left(1+e^{-| \omega | }\right)^2}
    \label{bmrootdens-nn}
\end{align}
\end{widetext}
The total number of Bethe roots is $ M=2+2\tilde\rho_d(0)=N,$
and hence, the total magnetization is
$S^z=N+1/2-M=\frac{1}{2}$,
which is the spin of the propagating spinon? Due to $SU(2)$ symmetry, there is another state with $S^z=-\frac{1}{2}$. The impurity is screened by this exponentially boundary-localized bound mode in this state.

The difference between the density given by Eq.\eqref{bmrootdens-nn} and Eq.\eqref{rootdens2strings-nn} is
\begin{align}
    \Delta_d\rho(\omega)&=-\frac{e^{\delta  | \omega | }+e^{-((\delta -2) | \omega | )}-2 e^{-\delta  | \omega | } \sinh (| \omega | )}{2 \left(e^{| \omega | }+1\right)^2}\nonumber\\
    &-\frac{\cos(\omega\theta)}{\left(e^{| \omega | }+1\right)^2}-\frac{\cos(\omega\lambda_c)\left(e^{-\frac{3}{2} (| \omega | )}+e^{-\frac{| \omega | }{2}}\right) }{ \left(2e^{-| \omega | }+e^{-2| \omega | }+1\right)},
\end{align}
where the first term is the contribution from the boundary string, the second term is from a hole at position $\theta$, and the third term is from the one-string located at $\lambda_c$.

The total energy of the boundary string (from the backflow effect and the bare energy) in the scaling limit is
\begin{equation}
    E_{\delta}=D\arctan(\frac{\cos(\pi\delta)}{\sinh(\pi b)})=m\cos(\pi \delta),
\end{equation}
 the energy of the hole is $E_\theta=m$, and the energy of the one-string vanishes in the thermodynamic limit. Thus, the energy of the excitation described by the root distribution Eq.\eqref{bmrootdens-nn} is 
 \begin{equation}
     E_{\theta,\delta}=m(1+\cos(\pi \delta),
 \end{equation}
which is zero at $\delta = 1$, attains a maximum value of $2m$ at $\delta = 2$, and is monotonically increasing for $1 < \delta < 2$. We shall refer to this excited state as Yu-Shiba-Rusinov (YSR), as the state is akin to the mid-gap Shiba states formed in the BCS superconductor in the presence of classical impurities \cite{Yu,Shiba,Rusinov}.

\begin{widetext}

\subsection{Density of states}
As mentioned earlier, the YSR regime has two unique low-energy states, the ground state with root distribution Eq.\eqref{rootdens2strings-nn} and the YSR state with root distribution Eq.\eqref{bmrootdens-nn}. Now, we focus on calculating the spinon density of states in both the ground state and the YSR state to determine whether the impurity is screened. The impurity density of states in the ground state (that is obtained from the root distribution Eq.\eqref{rootdens2strings-nn} is of the form
\begin{equation}
    \rho_{\mathrm{DOS}}^{\mathrm{imp}}(E)=\frac{\frac{2 \pi ^2 (\delta -1)}{\pi ^2 (\delta -1)^2+\cosh ^{-1}\left(\frac{E}{m}\right)^2}+\sum_{\nu=\pm}\psi ^{(0)}\left(\frac{1}{2} \left(\delta +\frac{\left(i \cosh ^{-1}\left(\frac{E}{m}\right)\right) \nu }{\pi }-1\right)\right)-\psi ^{(0)}\left(\frac{1}{2} \left(\delta +\frac{\left(i \cosh ^{-1}\left(\frac{E}{m}\right)\right) \nu }{\pi }\right)\right)}{2 \pi ^2 \sqrt{E^2-m^2}}
    \label{uns-state-dos}
\end{equation}
Notice that the integrated density of states contribution due to the impurity is
\begin{equation}
    \int_m^\infty\rho_{\mathrm{DOS}}^{\mathrm{imp}}(E)\mathrm{d}E=0,
\end{equation}
which demonstrates that the impurity is unscreened. 

In the YSR state, one one-string solution exists, the boundary string solution, and a hole on top of the 2-string sea. Thus, the total density of states due to the presence of the impurity obtained from the root density Eq.\eqref{bmrootdens-nn} now becomes

\begin{align}
    \rho_{\mathrm{DOS}}^\mathrm{imp}(E)&=-\frac{\text{sech}\left(\pi  \lambda_c -\cosh ^{-1}\left(\frac{E}{m}\right)\right)+\text{sech}\left(\pi  \lambda_c +\cosh ^{-1}\left(\frac{E}{m}\right)\right)}{4 \pi  \sqrt{E^2-m^2}}\nonumber\\
   & +\frac{2 m \left(\pi  (\delta -1) E \sinh (\pi -\pi  \delta )+\cosh (\pi -\pi  \delta ) \sqrt{E^2-m^2} \cosh ^{-1}\left(\frac{E}{m}\right)\right)}{\pi ^2 \sqrt{E^2-m^2} \left(-2 E^2+m^2 \cosh (2 \pi  (\delta -1))+m^2\right)}\nonumber\\
    &+\delta(E-E_\delta)
\end{align}
where the first term is the contribution from the one-string, the second term is the $\delta$ dependent impurity contribution, and finally, the third term is the direct contribution from the boundary string root. 

The impurity contribution to the integrated density of states
\begin{equation}
    \int_m^\infty\mathrm{d}E  \rho_{\mathrm{DOS}}^\mathrm{imp}=-1/2-1/4+1=1/4,
\end{equation}
which is positive, and hence, the impurity is screened by the bound mode. Both the one-string and the impurity term contribute negatively, but there is a sole positive contribution that comes from the boundary string root. Hence, this shows that this is a single mode of screening for impurity. 

In summary, the YSR phase arises when the strength of the bulk superconducting order is further increased, making the superconducting interaction comparable to or stronger than the Kondo coupling. In this regime, the impurity is no longer screened in the ground state, but a single-particle bound mode — akin to a Yu-Shiba-Rusinov state in a BCS superconductor — can screen it in an excited state. This midgap state has a finite mass, lying below the bulk superconducting gap, and provides an alternative mechanism for impurity screening distinct from the many-body Kondo cloud seen in the overscreened phase. The presence of a single-particle bound state localized at the edge reflects the competition between Kondo physics and superconductivity, where the latter dominates at low energies, preventing full Kondo screening.

\section{Local moment (unscreened) phase} When the impurity parameter $\delta>2$, the impurity is in the local moment phase as shown in the phase diagram Fig.\ref{fig:phasediag}. In this regime, the dressed energy boundary string solution Eq.\eqref{boundary-string-sol} vanishes in the thermodynamic limit \textit{i.e.} $E_\delta=0$,
as the energy due to the backflow exactly cancels its bare energy. Moreover, since $|\Im(\lambda_\alpha)|>\frac{3}{2}$, this solution becomes a ``wide string" \cite{destri1982analysis}, and hence, it is not possible to add this solution to the ground state without adding additional even number of massive holes which would make the energy of this state above the mass gap \cite{pasnoori2022rise}.  Thus, in this phase, the impurity is unscreened in the ground state. The ground state is made up of 2-strings whose root density is given in Eq.\eqref{rootdens2strings-nn}. The ground state is a two-fold degenerate state with $S^z=\frac{1}{2}$ where the spin contribution is due to there being an unscreened impurity. All the excited states can be constructed by adding an even number of spinons, bulk string solutions of various lengths, etc. The impurity density of states in the ground state is given by the analytic continuation of  Eq.\eqref{uns-state-dos} to this regime.

\section{Thermodynamics in the Kondo phase}\label{therm}
We shall now turn to the thermodynamics of the model, focusing on the Kondo phase. In the presence of boundary string solutions, the analysis of conventional thermodynamic Bethe Ansatz equations becomes more challenging. Thus, we set aside the analysis of impurity behavior in finite temperatures in other phases for future work.  As shown in Appendix~\ref{sec:TBA}, the thermodynamic Bethe Ansatz (TBA) equations \cite{TBAyang1969thermodynamics,TBAtakahashi1999thermodynamics,andrei1983solution} for the model in the Kondo phase are of the form
\begin{equation}
     \ln \eta_n(\lambda) = 
   -\frac{m}{T}\cosh(\pi \lambda)\delta_{n,2}+G\ln\left[ 1+\eta_{n+1} \right]+G\ln\left[ 1+\eta_{n-1} \right]
   \label{TBAeqn}
\end{equation}
To close these equations, we need to supply boundary conditions at $n\to \infty$, which gives a new relation
    \begin{equation}
    \lim_{n\rightarrow \infty} \left\{[n+1]\ln(1+\eta_n(\mu))-[n]\ln(1+\eta_{n+1}(\mu))\right\}=-\frac{H}{T}
\end{equation}
and also, we choose 
\begin{equation}
    \eta_0(\lambda)=0.
\end{equation}
Here, the functional \(\left[n\right]\) is defined as a convolution with the kernel \(K_n\), given by:  
\begin{equation}
    \left[n\right] g(\mu) \equiv K_n * g(\mu) = \int \dd \lambda \, K_n(\mu - \lambda) g(\lambda).
\end{equation}  
Additionally, the integral operator \(G\) is introduced and expressed as:  
\begin{equation}
    Gf(\lambda) = \int \dd \mu \, \frac{1}{2\cosh\pi (\lambda - \mu)} f(\mu).
\end{equation}

The spin part of the impurity free energy (see Appendix~\ref{sec:TBA} for details) is of the form
\begin{equation}
    \begin{aligned}
        \mathcal{F}_{\mathrm{imp}}&=\mathcal{F}_{\mathrm{imp}}^0-\frac{T}{2}\int \mathrm{d} \lambda\,\left(\frac{1}{2\cosh\pi (\lambda-d)}+\frac{1}{2\cosh\pi (\lambda+d)} \right)\ln(1+\eta_1(\lambda))
    \end{aligned}
    \label{freeeneg}
\end{equation}

We proceed to solve the coupled set of equations Eq.\eqref{TBAeqn} to determine $\eta_1(\lambda)$ for given $T, H$ and then from it compute the free energy from Eq.\eqref{freeeneg}.

\smallskip

For small temperatures, we can immediately solve Eq.\eqref{TBAeqn} to obtain
\begin{equation}
    \eta_2(\lambda)=e^{-\frac{m \cosh (\pi  \lambda )}{T}}.
    \label{eta2}
\end{equation}
We could further approximate $\eta_2(\lambda)\approx  e^{-\frac{m}{T}}$  upon realizing that most of the contribution only comes from $\lambda$ around $0$, we Taylor expanded the function around $\lambda=0$. 

With this, we can compute $\eta_1(\lambda)$ as
\begin{equation}
    \ln \eta_1(\lambda)=G \ln(1+\eta_2)=\frac{1}{2} \ln \left(e^{-\frac{m}{T}}+1\right).
\end{equation}
Or,
\begin{equation}
    \eta_1(\lambda)=\left(e^{-\frac{m}{T}}+1\right)^{1 /2}
\end{equation}

With this, using Eq.\eqref{freeeneg}, we compute the entropy as
\begin{equation}
    \mathcal{S}(T)=-\frac{\mathrm{d \mathcal{F}}}{\mathrm{d}T}=\frac{1}{4} \left(\frac{m}{T e^{m/T} \left(\sqrt{e^{-\frac{m}{T}}+1}+1\right)+T}+2 \log \left(\sqrt{e^{-\frac{m}{T}}+1}+1\right)\right)
    \label{entlo}
\end{equation}

We plot the following for various values of $m$. Notice that at $T\to 0$, the entropy $S=\ln\Omega$ is given by a non-integer $\Omega=\sqrt{2}$, which shows that the impurity exhibits non-Fermi liquid behavior. This is consistent with the fact that the spin-1 fermions in the bulk overscreen the spin-$\frac{1}{2}$ impurity in the boundary. 

Keeping all order in Eq.\eqref{eta2}, we can compute $\eta_1(\lambda)$ and hence the free energy and the entropy numerically to see the $d$ dependent contribution which is of the reform $S(T)=S(m/T,d)$. Below, we plot the result of such a calculation, showing the values of $S(m/T,d)$ for various representative values of $m$ and $d$, which shows that irrespective of $m$ and $d$, the zero-temperature of entropy is $\ln\sqrt{2}$ which asymptotically approaches to $\ln 2$ as $T\to \infty$.
\begin{figure}[H]
    \centering
    \includegraphics[width=0.75\linewidth]{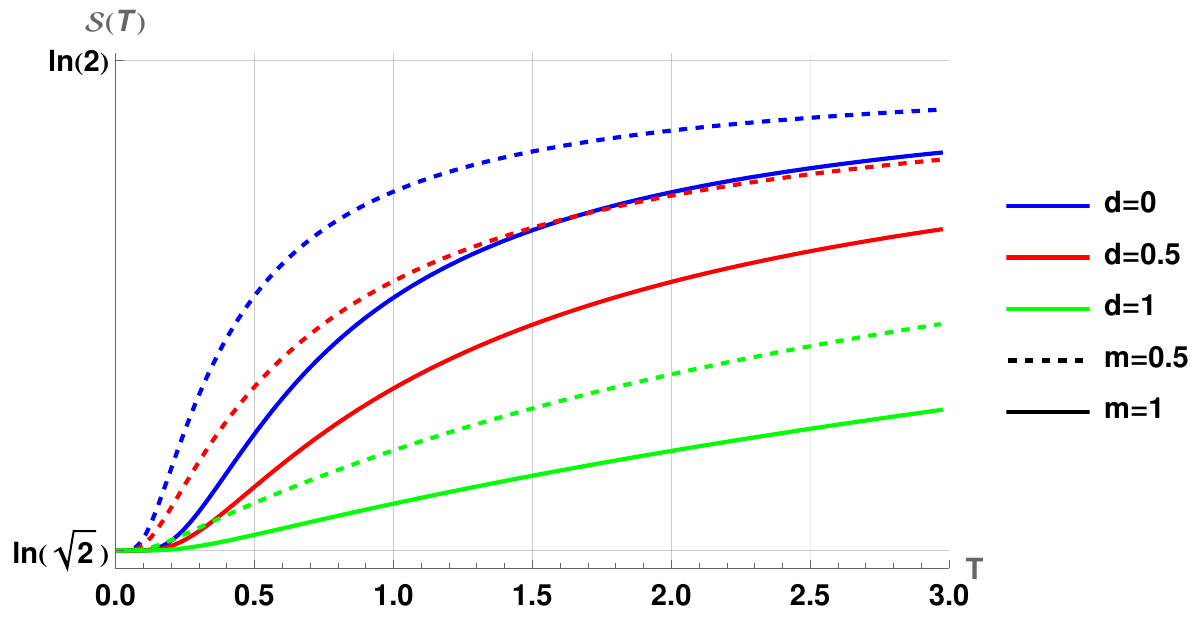}
    \caption{Impurity entropy for $m=0.5$ (shown in dashed lines) and for $m=1$ (shown in solid lines) obtained by numerically solving Eq.\eqref{TBAeqn} using Eq.\eqref{eta2} as the driving term. The plot shows that irrespective of the RG invariant quantity $d$, the impurity entropy is $\ln \sqrt{2}$ at $T\to 0$, and it smoothly crosses over to $\ln 2$ as $T\to \infty$.}
    \label{fig:ddep}
\end{figure}
\end{widetext}
Fig.~\ref{fig:ddep} shows that irrespective of values of real $d$ in the Kondo phase, the impurity entropy starts from $\ln(\sqrt{2})$ at $T\to 0$ and asymptotes to $\ln (2)$ at high temperature. This shows that the impurity is overscreened at low temperatures and is asymptotically free at high temperatures.

\section{Discussion}

We considered the behavior of a single spin-\(\frac{1}{2}\) magnetic impurity coupled to the edge of a one-dimensional spin-1 superconductor. The system is described by an integrable Gross-Neveu-like \(O(3)\) invariant model with attractive interactions between fermions of opposite chirality. A key feature of this model is the competition between the Kondo effect and superconductivity, leading to a dynamically generated mass gap \(2m\) in the bulk and a rich phase structure for the impurity at the boundary.

Regardless of the initial coupling strength, the bulk interaction parameter \( g \) always flows to strong coupling under the renormalization group (RG) flow, ensuring the formation of a superconducting gap. However, the fate of the impurity is governed by the relative strength of the Kondo coupling \(J\) and the bulk interaction \(g\), leading to four distinct impurity phases.

When the Kondo coupling is strong (\(J \gg g\)), the impurity undergoes \textit{many-body overscreening} by a cloud of bulk spin-1 excitations. This results in an \textit{overscreened Kondo phase}, similar to the two-channel Kondo effect, where the impurity entropy at low temperatures remains nonzero at \(\frac{1}{2} \ln 2\), signaling a non-Fermi liquid fixed point. The system dynamically generates a Kondo energy scale \( T_K \), which governs impurity screening and is distinct from the dynamically generated bulk superconducting gap \( \Delta = 2m \).

In contrast, when the bulk superconducting interaction dominates (\(J \ll g\)), the system enters a \textit{local moment (unscreened) phase}, where the impurity cannot be screened at any energy scale. Here, the impurity remains decoupled from the bulk due to the strong superconducting order, behaving as a free spin-\(\frac{1}{2}\) at the boundary.

For intermediate values of \(J\) and \(g\), two novel impurity phases emerge due to the delicate balance between Kondo screening and superconductivity. In the \textit{zero mode phase}, the impurity is still overscreened by the many-body Kondo cloud, but an additional \textit{zero-energy boundary excitation} appears, described by a purely imaginary Bethe root. This boundary excitation leads to an \textit{exact degeneracy between the ground state and an excited state} in the thermodynamic limit, making this phase distinct from the conventional overscreened Kondo regime.

Further increasing the superconducting order relative to the Kondo coupling leads to the \textit{YSR phase}, where the impurity is \textit{unscreened in the ground state} but can be screened in an excited state by a \textit{single-particle bound mode} localized at the edge. This bound mode, akin to a \textit{Yu-Shiba-Rusinov (YSR) state} in a BCS superconductor, lies inside the superconducting gap but below the bulk excitation continuum. Unlike the Kondo phase, where the impurity is screened by a collective many-body effect, the YSR phase allows for a screening mechanism via an individual bound state, demonstrating an alternative way in which impurity physics manifests in superconducting systems.

The transitions between these phases are governed by an \textit{RG-invariant parameter} \( d(J, g) \), which encodes the running of both the bulk and boundary couplings. This parameter determines the stability of the impurity phases and characterizes the crossover between different regimes. The competition between Kondo physics and superconductivity thus leads to a rich boundary phase diagram, illustrating how impurity behavior is modified in strongly correlated superconducting systems.

While our current results are derived in the thermodynamic limit, it would be valuable to extend the analysis using finite-size scaling to enable direct comparisons with perturbed CFT and NRG results. Such an approach would reveal the differences in the impurity's thermodynamic and dynamical behavior across these four phases, offering deeper insights into the interplay between impurity and bulk properties.

\section*{Acknowledgement}
Stimulating discussions with Colin Rylands and Anirvan Sengupta are greatly appreciated.

\bibliography{ref}

\widetext

\appendix

\section{Hamiltonian as $SU(2)$ spin$-1$ Gross-Neveu model with a boundary impurity}\label{su2level2}

Notice that upon performing the unitary rotation
\begin{equation}
    S^a=U^\dagger \tilde \tau^a U
\end{equation}
where
\begin{equation}
    U=\left(
\begin{array}{ccc}
 -\frac{i}{\sqrt{2}} & 0 & \frac{i}{\sqrt{2}} \\
 \frac{1}{\sqrt{2}} & 0 & \frac{1}{\sqrt{2}} \\
 0 & i & 0 \\
\end{array}
\right),
\end{equation}
the three generators in written in Eq.\eqref{rep2} become such that the third component is diagonal \textit{i.e.}
\begin{equation}
   S^x=\frac{1}{\sqrt{2}}\left(
\begin{array}{ccc}
 0 & 1 & 0 \\
 1 & 0 & 1 \\
 0 & 1 & 0 \\
\end{array}
\right)\ \quad S^y=\frac{1}{\sqrt{2} i}\left(
\begin{array}{ccc}
 0 & 1 & 0 \\
 -1 & 0 & 1 \\
 0 & -1 & 0 \\
\end{array}\right) \quad
 S^z=\left(
\begin{array}{ccc}
 1 & 0 & 0 \\
 0 & 0 & 0 \\
 0 & 0 & -1 \\
\end{array}\right).
\label{rep1}
\end{equation}

This is the three-dimensional spin-1 representation of $SU(2)$. In terms of these matrices, the Hamiltonian Eq.\eqref{modelham} becomes
 \begin{align}
    H&=\int_{-L/2}^0\mathrm{d}x\left[-i(\psi^\dagger_{r+}\partial_x \psi_{r+}-\psi^\dagger_{r-}\partial_x \psi_{r-})+2g \psi^\dagger_{r+}\psi^\dagger_{s-}[\vec S\cdot \vec S+I]_{rs}^{uv} \psi^\dagger_{u-}\psi^\dagger_{v+}+2j \psi^\dagger_{r,-}(0)\vec S_{rs}\psi_{r,+}(0)\cdot\vec \sigma\right],
    \label{modelham1}
\end{align}

This is the $SU(2)$ spin-$1$ Gross-Neveu model. In this representation, it is easier to obtain the Bethe equations of the model and also to label the multiparticle eigenstates using the good quantum number $S^z$, the $ z$-component of the total spin. We shall now construct the $ N$-particle eigenstates.

\section{N-particle eigenstates}\label{Nparteig-sec}
   Notice that since $N$ is a good quantum number, we can construct the eigenstates by examining each particle sector independently.
   
   \subsection{One particle sector}
   Starting with $N=1$ fermion and the single impurity, the wavefunction can be expressed as
   \begin{equation}
       \ket{k}=\sum_{a_j=\uparrow \downarrow,\sigma=\pm}\int_{-L}^0 \mathrm{d}x e^{i\sigma k x}A^\sigma_{a_1 a_0}\psi_{\sigma,a_1}^\dagger (x) \ket{0}\otimes \ket{a_0},
   \end{equation}
   where $\ket{0}$ is the vacuum $\psi_{\sigma,a_j}\ket{0}=0$. Here $A^\sigma_{a_j a_0}$ are the amplitudes for the $j^{\mathrm{th}}$ itinerant fermion with chirality index $\sigma$ and spin $a_j$ scattering off the localized impurity carrying spin $a_0$. Applying the Hamiltonian to the state $\ket{k}$, Schrodinger's equation fixes these amplitudes in the following form
   \begin{equation}
       A^-_{a_ja_0}={S^{j0}}_{a_ja_0}^{b_jb_0} A^+_{b_jb_0},
   \end{equation}
   where the scattering fermion-impurity bare scattering matrix is given by
   \begin{equation}
       S^{j0}=\frac{i I^{j0}+J \vec S^j\cdot \vec \sigma^0}{i I^{j0}-J\vec S^j\cdot \vec \sigma^0}=e^{i\phi}\frac{I^{j0}\left( 1-i\frac{c}{2}\right)-ic \vec S^j\cdot \vec \sigma^0}{1-\frac{3 i c}{2}},
\label{imp-part-smat}   \end{equation}
   where we suppressed the spin indices, and here
   \begin{equation}
       c=\frac{2 J}{1-2 J^2} \quad\quad \text{and}\quad\quad \phi=2 \tan ^{-1}(J),
   \end{equation}
   and $I^{j0}$ is the identity operator; throughout, we use the convention that superscripts denote the spaces in which the operators act non-trivially. Since in one particle sector, the only interaction involves the fermion interacting with the impurity and changing its chirality, the identification of the impurity-electron S-matrix $S^{j0}$ completes the construction of the eigenstate $\ket{k}$ which has eigenvalue $E=k$.

\subsection{Two particle sector}
We shall now consider the two-particle sector where the bulk interaction with coupling strength $g$ in the Hamiltonian Eq.\eqref{modelham1} comes into play. As usual, we shall write the wave function as the sum of plane waves with different amplitudes in different regions that are separated by the ordering of the particles \textit{i.e.}
\begin{equation}
\ket{k_i,k_j}=\sum_{\sigma,a}\int_{-L}^0 \mathrm{d}^2x F_{a_ia_ja_0}^{\sigma_i\sigma_j}(x_i,x_j)e^{i k_i \sigma _i x_i+i k_j \sigma _j x_j}\psi_{\sigma_ia_i}^\dagger(x_i)\psi_{\sigma_j a_j}^\dagger(x_j)\ket{0}\otimes \ket{a_0},
\label{2partES}
\end{equation}
where the sum is taken over all spin and chirality configurations and $F_{a_ia_ja_0}^{\sigma_i\sigma_j}(x_i,x_j)$ is the two-particle wave function, which is written in terms of the scattering amplitudes in different sectors depending on the ordering of the particles as
\begin{equation}
   F_{a_ia_ja_0}^{\sigma_i\sigma_j}(x_i,x_j)=A_{a_ia_ja_0}^{\sigma_i\sigma_j}[ij]\theta(x_j-x_i)+ A_{a_ia_ja_0}^{\sigma_i\sigma_j}[ji]\theta(x_i-x_j).
\end{equation}
Here, $A_{a_ia_ja_0}^{\sigma_i\sigma_j}[\mathcal{C}]$ are the amplitudes for configurations with specified spin $a_j$ and chirality $\sigma_j$ as well as the ordering of particles in the configuration space, for example, $\mathcal{C}=ij$ refers to the configuration where $i^\mathrm{th}$ is to the left of the $j^{\mathrm{th}}$ particle and $\mathcal{C}=ji$ refers to the configuration where the particles are exchanged.

Upon applying the Hamiltonian Eq.\eqref{modelham1} on the state Eq.\eqref{2partES}, we find that it is an eigenstate with eigen energy 
\begin{equation}
    E=k_i+k_j,
\end{equation}
if the various amplitudes are related to each other via different scattering matrices introduced below. 

When the right-most particle scatters off the impurity, its chirality is changed. In that case, the amplitudes before and after scattering are related by the impurity-particle $S-$matrix obtained in Eq.\eqref{imp-part-smat}. Suppressing the spin indices, we write the relations as
\begin{align}
    A^{\sigma_i-}[ij]&=S^{j0}A^{\sigma_i-}\\
    A^{-\sigma_j}[ji]&=S^{i0}A^{-\sigma_j}.
\end{align}

The exchange interaction gives rise to two different kinds of scattering matrices denoted by $S^{ij}$ and $W^{ij}$. Here, $S^{ij}$ relates the scattering amplitudes when the particles of opposite chirality are exchanged:
\begin{align}
    A^{+-}[ji]&=S^{ij}A^{+-}[ij]\\
    A^{-+}[ij]&=S^{ij}A^{-+}[ji]
\end{align}
where $S^{ij}$ acts on the color spaces of $i^{\mathrm{th}}$ and $j^{\mathrm{th}}$ particles whose explicit form is
\begin{equation}
    S^{ij}=e^{i\zeta}\left(\frac{i-2 b}{i+2 b}I-\frac{i}{b-i}\vec S \cdot\vec S-\frac{1}{(i-b) (i-2 b)}(\vec S\cdot\vec S)^2 \right),
\end{equation}
where 
\begin{equation}
    b=\frac{1}{2g} \quad\quad \text{and}\quad\quad \zeta=-2\tan^{-1}{2g}.
\end{equation}
Likewise, $W^{ij}$ relates the scattering amplitudes when the particles of the same chirality are exchanged:
\begin{align}
    A^{--}[ji]&=W^{ij}A^{+-}[ij]\\
    A^{++}[ij]&=W^{ij}A^{-+}[ji].
\end{align}

Unlike the scattering matrices $S^{ij}$, the scattering matrices $W^{ij}$ are not dictated by the Hamiltonian, as there is no interaction between the particles of the same chirality. Their matrices are fixed via a consistency relation called the reflection equation
\begin{equation}
    S^{i0}S^{ij}S^{j0}W^{ij}=W^{ij}S^{j0}S^{ij}S^{i0},
\end{equation}
which is readily satisfied if we take $W^{ij}=P^{ij}$

Having obtained these various scattering matrices, we have now completed the diagonalization of the Hamiltonian in the two-particle sector.

\subsection{N-particle sector}
Generalization to the $N-$particle sector is now fairly straightforward. We construct eigenstates
with eigenvalues 
\begin{equation}
    E = \sum_{j=1}^{N} k_j.
\end{equation}
of the form
\begin{equation}
| \{ k_j \} \rangle = \sum_{Q, \vec{a}, \vec{\sigma}} \int \theta(x_Q) A_{\vec{a}}^{\vec{\sigma}}[Q] \prod_{j=1}^{N} e^{i \sigma_j k_j x_j} \psi_{a_j \sigma_j}^{\dagger}(x_j) | 0 \rangle \otimes | a_0 \rangle,
\label{NpartES}
\end{equation}
whereas before the sum is taken over all spin and chirality configurations that are specified by $\vec a=\{a_1,\cdots,a_N,a_0\}$ and $\vec \sigma=\{\sigma_1,\cdots,\sigma_N,\sigma_0\}$ respectively as well as the different orderings of the particles in the configuration space which correspond to elements of the symmetric group $Q \in S_N$. Furthermore, $\theta(x_Q)$ represents the Heaviside function, which is nonzero only for a specific ordering. In the $N=1, 2$ sectors, the amplitudes $A_{\vec{a}}^{\vec{\sigma}}[Q]$ are interconnected by various $S$ matrices in the same manner as previously described. Notably, the amplitudes differing by the chirality change of the rightmost particle $j$ scattering off the impurity are linked by the impurity-particle $S$ matrix, $S^{j0}$. Those that differ by the order exchange of particles with opposite or identical chiralities are related by $S^{ij}$ and $W^{ij}$, respectively. The validity of this construction is ensured by the $S-$ matrices satisfying the reflection and Yang-Baxter equations:
\begin{align}
W^{jk} W^{ik} W^{ij} &= W^{ij} W^{ik} W^{jk}, \label{Wrel}\\
S^{jk} S^{ik} W^{ij} &= W^{ij} S^{ik} S^{jk}, \label{SWrel}\\
S^{j0} S^{ij} S^{i0} W^{ij} &= W^{ij} S^{i0} S^{ij} S^{j0} \label{YEBrel},
\end{align}
where $W^{ij} = P^{ij}$ with the superscripts indicating the color space of the particles on which the operators act non-trivially, as previously mentioned. These four relationships are sufficient to construct a consistent $N$-particle eigenstate.

\section{Integrability of the model}\label{integrability}
We shall proceed to show that this model is integrable.  We shall first impose open boundary conditions
\begin{equation}
    \psi_+(0)=-\psi_-(0) \quad \text{and}\quad \psi_+(-L)=-\psi_-(-L).
    \label{openBC}
\end{equation}
Subjecting the $N-$particle eigenstate Eq.\eqref{NpartES} to the boundary condition Eq.\eqref{openBC} at $x=-L$, we obtain the quantization conditions for the quasimomenta
\begin{equation}
    e^{-2ik_jL}A_{\vec{a}}^{\vec{\sigma}}[Q] = \left(Z_{j}\right)_{ \vec{a} \vec{a}'}^{\vec{\sigma}, \vec{\sigma}'}A_{\vec{a}'}^{\vec{\sigma}'}[Q],
    \label{quantization-cond}
\end{equation}
where the operator $Z_j$ refers to the transfer matrix of the $j^{\mathrm{th}}$ particle which takes the form
\begin{equation}
    Z_j = W^{jj-1}\cdot W^{j1}S^{jA}S^{j1}\cdot S^{jN}S^{jB}W^{jN}\cdot W^{jj+1}.
    \label{tmateqn}
\end{equation}
This transfer matrix is an operator that `transports' the $j^{\mathrm{th}}$ particle from one end of the system to the other and returns it to the original place, gathering $S$ matrix factors along the way as it scatters off the remaining $N-1$ particles, first as a right mover and then as a left mover once its chirality changes as it scatters off the impurity. Using the relations Eq.\eqref{Wrel}-Eq.\eqref{YEBrel}, it is easy to prove that the transfer matrices $Z_j$ and $Z_k$ for any $j,k$ commute \textit{i.e.}  $[Z_j,Z_k]=0$, which shows that all of these transfer matrices can be diagonalized simultaneously. In order to diagonalize these transfer matrices, we use the functional Bethe Ansatz method \cite{sklyanin1990functional,wang2015off} and fusion hierarchy \cite{wang2015off,mezincescu1992fusion,BABUJIAN1983317,Cao_2015}. In order to proceed with this formalism, we need to embed the bare S-matrices derived previously into a continuous framework. This means determining a spectral parameter-dependent R-matrix $R(\lambda)$ and a boundary reflection matrix $K(\lambda)$ so that, for particular values of the spectral parameter $\lambda$, we can obtain all the previously derived bare S-matrices for the model.

Choosing the $R-$matrix of 19-vertex model 
\begin{equation}
    R^{ij}(\lambda)=\frac{\lambda+i}{\lambda-i}I^{ij}-\frac{2i}{\lambda-2i}\vec S\cdot \vec S - \frac{2}{(\lambda-i)(\lambda-2i)}\left(\vec S\cdot \vec S \right)^2.
\end{equation}
we obtain the two bulk S-matrices as $R^{ij}(0)=W^{ij}$ and $R^{ij}(2b)=S_{ij}$ upto the unimportant phase factor $e^{i\zeta}$ which we shall ignore. For the boundary matrix, we shall choose $K^{j0}(\lambda)=r^{j0}(\lambda+d)r^{j0}(\lambda-d)$ with $d=\sqrt{b^2-\frac{2 b}{c}-\frac{9}{4}}$, such that $K^{j0}(b)=S^{j0}$.
Here,
\begin{equation}
    r^{ij}(u)=\frac{1}{u-3i/2}\left(I^{ij}\left(u-\frac{i}{2}\right)-i\vec S^i\cdot\vec \sigma^j \right).
\end{equation}

It is often useful to write the $R-$matrix $R^{ij}$ in terms of the spin projection operators to compare with other literature \cite{wang2015off,BABUJIAN1983317}
\begin{equation}
    R_{ij}(u)= \frac{(u+i )(u+2 i )}{(u-i) (u-2i)} \left(\mathcal P_{ij}^{(0)}+\frac{u-i}{u+i }\mathcal P_{ij}^{(1)}+\frac{ (u-i) (u-2i)}{(u+i ) (u+2 i )}\mathcal P_{ij}^{(2)}\right),
\end{equation}
where  $\mathcal P_{i,j}^{(\ell)}$ is a projector defined in the tensor space of two spin$-s$, which projects the tensor space into the irreducible subspace of spin$-\ell$. These projectors can succinctly be written as
\begin{equation}
P^{(j)}=\prod_{\substack{\ell=0 \\ \ell \neq j}}^{2 s} \frac{x-x_\ell}{x_j-x_\ell},
\end{equation}
where $x=\vec{S}_1 \cdot\vec{S}_2$ and $x_\ell=\frac{1}{2}[\ell(\ell+1)-2 s(s+1)]$.

The transfer matrix \(Z_1\) is related to the Monodromy matrix \(\Xi_s^{A}(\lambda)\) as \(Z_1 = \tau(b) = \text{Tr}_{A} \Xi^{A}(b)\), where
\begin{equation}
\Xi^{A}(\lambda) = R^{A1}(\lambda + b) \cdots R^{AN}(\lambda + b) r^{A0}(\lambda + d) r^{A0}(\lambda - d)
\times R^{AN}(\lambda - b) \cdots R^{A1}(\lambda - b).
\label{monodromy}
\end{equation}
Here, \(A\) represents an auxiliary space which is taken to be that of a spin-$1$ particle, and \(\text{Tr}_{A}\) represents the trace over the auxiliary space. Using the properties of the \(R\)-matrices and $r-$matrices, one can prove that \([\tau(\lambda), \tau(\mu)] = 0\) and by expanding \(\tau(\mu)\) in powers of \(\mu\), obtain an infinite set of conserved charges which guarantees integrability. By following the functional Bethe Ansatz approach and $T-Q$ relation \cite{wang2015off}, we shall now obtain the Bethe equations.

\subsection{Bethe Ansatz Equations}

To make the notations similar to \cite{BABUJIAN1983317,wang2015off}, let us introduce the two $R-$matrices
\begin{equation}\label{sigma s R}
R_{\sigma s}^{ij}(u)=\left(u+\frac{\eta}{2}\right)I^{i,j}+\eta \;\vec{\sigma}_i \cdot \vec{S}_j
\end{equation}
and 
\begin{equation}\label{s R}
R_{s}^{ij}(u)=\prod_j^{2s}(u-j\eta)\sum_{l=0}^{2 s} \prod_{k=1}^\ell \frac{u+k \eta}{u-k \eta} P^{(\ell)}_{ij},
\end{equation}
where the former represents an $R$-matrix between particles with spins 1/2 and $s$, and the latter represents an $R$-matrix between particles with spin $s$.
Our $R$ and $r$ matrices are related to these by
\begin{equation}
    R_{s}^{ij}(u)=(u+\eta)(u+2\eta)R_{ij}(u) \quad \text{and}\quad R_{\sigma s}^{ij}(u)=\left(u+\frac32\eta\right)r_{ij}(u),
\end{equation}
when the crossing parameter is taken to be $\eta=-i$ and $s=1$.

Let us also introduce the $R$ matrix when both particles have spin-$\frac{1}{2}$:
\begin{equation}
     R_{\sigma}^{ij}(u)=u+\frac{\eta}{2}(1+\vec \sigma_i\cdot\vec \sigma_j).
\end{equation}
In the previous section, we introduced a monodromy matrix Eq.\eqref{monodromy} $\Xi_s^A(u)$, where we worked with an auxiliary particle of spin-$1$ that was subsequently traced out, yielding the transfer-matrix $t_s(u)$. It's worth noting that, in general, since the auxiliary particle is traced out, it can be considered to possess any arbitrary spin $j$.

In this section, we shall write the monodromy matrix with auxiliary space $A$ taken to be that of a spin-$\frac{1}{2}$ particle. Then the monodromy matrix can be written as
\begin{equation}
   \Xi^A_\sigma(\lambda)= R_{\sigma s}^{A 1}(\lambda+b)  \cdots R_{\sigma s}^{AN}(\lambda+b) R_{\sigma }^{A0} (\lambda+d) R_{\sigma}^{A0} (\lambda-d) R_{\sigma s}^{AN}(\lambda-b) \cdots R_{\sigma s}^{A1}(\lambda-b),
   \label{mon-a-spin12}
\end{equation}
and the resultant transfer matrix is $t_\sigma(b)=\mathrm{Tr}_A\Xi_\sigma^A(b)$.

Now, we notice that
\begin{equation}
   R_{\sigma s}^{ij}(u)R_{\sigma s}^{ij}(-u) = -\left(u+\left(\frac{1}{2}+s\right)\eta\right)\left(u-\left(\frac{1}{2}+s\right)\eta\right)  I^{ij},
\end{equation}
and
\begin{equation}
     R_{\sigma}^{ij}(u)R_{\sigma}^{ij}(-u) = -\left(u+\eta\right)\left(u-\eta\right)  I^{ij}.
\end{equation}

Thus, the eigenvalues of the operator $\tau(\lambda)$ satisfy the Baxter's $T-Q$ relation of the form
\begin{equation}
    \Lambda(\lambda)=a(\lambda) \frac{Q(\lambda-\eta)}{Q(\lambda)}+\frac{Q(\lambda+\eta)}{Q(\lambda)} d(\lambda),
    \label{eqnlamlam}
\end{equation}
where the Q-function is given by
\begin{equation}
Q(\lambda)=\prod_{\ell=1}^M (\lambda-\lambda_\ell	)(\lambda+\lambda_\ell+\eta)
\end{equation}
and
\begin{align}
    a(\lambda)&=\frac{2(\lambda+\eta)}{2\lambda+\eta}\left(\left(\lambda+\left(\frac{1}{2}+s\right)\eta\right)^2-b^2\right)^{N}((\lambda+\eta)^2-d^2)\\
    d(\lambda)&=a(-\lambda-\eta)=\frac{2\lambda }{\eta +2 \lambda }  \left( \left(\lambda + \left(\frac{1}{2}-s\right)\eta \right)^2-b^2\right)^2 (\lambda^2-d^2).
\end{align}

Regularity of the T-Q equation gives the Bethe Ansatz equations
\begin{equation}
\frac{\lambda_j +\eta}{\lambda_j }\left(\frac{\lambda_j +b+\left(\frac{1}{2}+s\right)\eta}{\lambda_j +b+\left(\frac{1}{2}-s\right)\eta}\right)^N\left(\frac{\lambda_j -b+\left(\frac{1}{2}+s\right)\eta}{\lambda_j -b+\left(\frac{1}{2}-s\right)\eta}\right)^N\frac{\lambda_j +d+\eta}{\lambda_j +d}\frac{\lambda_j -d+\eta}{\lambda_j -d}=-\prod_{\ell=1}^M\frac{(\lambda_j-\lambda_\ell+\eta)(\lambda_j+\lambda_\ell+2\eta)}{(\lambda_j-\lambda_\ell-\eta)(\lambda_j+\lambda_\ell)}.
\label{prebae}
\end{equation}
Upon changing the variable $\lambda_j\to \lambda_j-\frac{\eta}{2}$ and recalling that $\eta=-i$ and $s=1$, we arrive at the Bethe Ansatz equations written in the main text
\begin{equation}
  \left(\frac{\lambda_j+b+i}{\lambda_j+b-i} \right)^N \left(\frac{\lambda_j-b+i}{\lambda_j-b-i} \right)^N \frac{\lambda_j+d+\frac{i}{2}}{\lambda_j+d-\frac{i}{2}} \frac{\lambda_j-d+\frac{i}{2}}{\lambda_j-d-\frac{i}{2}}=\prod_{\ell\neq j}^M\frac{\lambda_j-\lambda_\ell+i}{\lambda_j-\lambda_\ell-i}\frac{\lambda_j+\lambda_\ell+i}{\lambda_j+\lambda_\ell-i}.
\end{equation}

However, to derive the relation between these Bethe roots and the transfer matrix, we need to consider the transfer matrix defined with the auxiliary space of a spin-1 particle. 

Writing out the monodromy matrix with auxiliary space taken to be that of a spin-1 particle
\begin{equation}
   \Xi^A_\sigma(\lambda)= R_{ s}^{A 1}(\lambda+b)  \cdots R_{ s}^{AN}(\lambda+b) R_{s\sigma }^{A0} (\lambda+d) R_{s\sigma}^{A0} (\lambda-d) R_{s}^{AN}(\lambda-b) \cdots R_{ s}^{A1}(\lambda-b),
   \label{mon-a-spin12-}
\end{equation}
and the resultant transfer matrix is 
\begin{equation}
    t_s(b)=\mathrm{Tr}_A\Xi_s^A(b).    \label{tmatspin1}
\end{equation}
Following \cite{wang2015off}, we find the transfer matrices defined with auxiliary space of spin$-1$ and spin-$\frac{1}{2}$ are related as
\begin{equation}
    t_s(u) = \frac{u (\eta +u)}{\left(\frac{\eta }{2}+u\right) \left(\frac{3 \eta }{2}+u\right) \left(u-d+\frac{\eta }{2}\right) \left(u+d+\frac{\eta }{2}\right)}\left[t_\sigma\left(u + \frac{\eta}{2}\right) t_\sigma\left(u - \frac{\eta}{2}\right) - \delta_s\left(u + \frac{\eta}{2}\right)\right],
    \label{fusedrel}
\end{equation}
where the quantum determinant $\delta_s\left(u\right)$ is given as
\begin{equation}
    \delta_s(u)=a(u)d(u-\eta).
\end{equation}

Comparing this transfer matrix Eq.\eqref{tmatspin1} with the transfer matrix built out of the bare $S-$matrix Eq.\eqref{tmateqn} and the quantization condition Eq.\eqref{quantization-cond}, we arrive at the relation
\begin{equation}
    e^{-2ik_jL}=\frac{t_s(b)}{2^N \left(\eta ^2\right)^N \left(b-d+\frac{3 \eta }{2}\right) \left(b+d+\frac{3 \eta }{2}\right) ((2 b+\eta ) (2 b+2 \eta ))^N}.
\end{equation}

Computing $t_s(b)$ using the fusion relation Eq.\eqref{fusedrel}, we arrive at the Bethe Ansaz equation
\begin{equation}
    e^{-2ik_j L}=\frac{Q\left( b-\frac{3\eta}{2}\right)}{Q\left( b+\frac{\eta}{2}\right)}.
\end{equation}
Upon performing the transformation $\lambda_\alpha\to \lambda_\alpha-\frac{\eta}{2}$ and plugging in $\eta=-i$, we obtain
\begin{equation}
    e^{-2ik_j L}=\prod_{\alpha=1}^M \left(\frac{ b+\lambda_\alpha+i}{ b+\lambda_\alpha-i} \right) \left(\frac{ b-\lambda_\alpha+i}{ b-\lambda_\alpha-i} \right)
    \label{CBAEapp}
\end{equation}
reported in the main text.

\section{Alternative derivation of the Bethe Ansatz equations}\label{altBAE}
One of us studied the problem in the absence of impurity with periodic boundary conditions previously~\cite{andrei1984dynamical}. There, the Bethe equations were obtained by performing the symmetric fusion at the level of the Bethe equations by noticing that the spin-$1$ model can be constructed by taking the two copies of spin-$\frac12$ models and performing symmetric fusion. Generalizing the detailed construction outlined in \cite{andrei1984dynamical}  for the case of periodic boundary conditions to our case with open boundary conditions and impurity, we shall derive the Bethe equations for our present model in this section. 

First plugging in $s=1/2$ and $\eta=-i$ in Eq.\eqref{prebae} and changing variable $\lambda_j\to \lambda_j+\frac{i}{2}$, we write the Bethe equations of spin-$\frac12~SU(2)$ Gross-Neveu model with spin-$\frac12$ impurity  as
\begin{equation}
  \left(\frac{\lambda_j+b+\frac{i}{2}}{\lambda_j+b-\frac{i}{2}} \right)^N \left(\frac{\lambda_j-b+\frac{i}{2}}{\lambda_j-b-\frac{i}{2}} \right)^N \frac{\lambda_j+d+\frac{i}{2}}{\lambda_j+d-\frac{i}{2}} \frac{\lambda_j-d+\frac{i}{2}}{\lambda_j-d-\frac{i}{2}}=\prod_{\ell\neq j}^M\frac{\lambda_j-\lambda_\ell+i}{\lambda_j-\lambda_\ell-i}\frac{\lambda_j+\lambda_\ell+i}{\lambda_j+\lambda_\ell-i}.
  \label{BAEspin1/2Wspin1/2}
\end{equation}

Now, using the fact that the eigenvalues of the bare transfer matrix $e^{2ik_jL}$ are related to the eigenvalues of the transfer matrix Eq.\eqref{eqnlamlam}, following the same routine described above, we arrive at the Bethe equation
\begin{equation}
    Z_j=e^{-2ikjL}=\frac{Q(b-\eta)}{Q(b)},
\end{equation}
and upon performing the change of variable $\lambda_j\to \lambda_j+\frac{i}{2}$, we arrive at the Bethe equation
\begin{equation}
    e^{-2ik_j L}=\prod_{\alpha=1}^M \left(\frac{ b+\lambda_\alpha+\frac{i}{2}}{ b+\lambda_\alpha-\frac{i}{2}} \right) \left(\frac{ b-\lambda_\alpha+\frac{i}{2}}{ b-\lambda_\alpha-\frac{i}{2}} \right).
    \label{quasimomrel}
\end{equation}
These Bethe equations were obtained in \cite{pasnoori2020kondo} using the boundary algebraic Bethe Ansatz method.

We shall now take 2 copies of these spin-$\frac{1}{2}$ Gross-Neveu model, one with impurity and one without impurity, and write the Bethe equations as
    \begin{equation}
   \frac{\lambda_j+d+\frac{i}{2}}{\lambda_j+d-\frac{i}{2}} \frac{\lambda_j-d+\frac{i}{2}}{\lambda_j-d-\frac{i}{2}}\prod_{j=1}^{2N} \left(\frac{\lambda_\alpha-\mu_j+\frac{i}{2}}{\lambda_\alpha-\mu_j-\frac{i}{2}}\right)\left(\frac{\lambda_\alpha+\mu_j+\frac{i}{2}}{\lambda_\alpha+\mu_j-\frac{i}{2}}\right)=\prod_{\beta\neq\alpha}^M \frac{\lambda_\alpha-\lambda_\beta+i}{\lambda_\alpha-\lambda_\beta-i}\frac{\lambda_\alpha+\lambda_\beta+i}{\lambda_\alpha+\lambda_\beta-i}.
   \label{combinedeqn2}
\end{equation}
and the energy equation as
\begin{equation}
    e^{-2ik_j L}=\prod_{\alpha=1}^{2M} \left(\frac{ \mu_j+\lambda_\alpha+\frac{i}{2}}{ \mu_j+\lambda_\alpha-\frac{i}{2}} \right) \left(\frac{ \mu_j-\lambda_\alpha+\frac{i}{2}}{ \mu_j-\lambda_\alpha-\frac{i}{2}} \right).
    \label{quasimomrell}
\end{equation}
Here, we introduced the inhomogeneity parameters $\mu_j$ only in the bulk to proceed with the fusion procedure. To complete the symmetric fusion only in the bulk, we put 
\begin{equation}
    \mu_{2j-1}=b-\frac{i}{2} \quad \text{and}\quad \mu_{2j}=b+\frac{i}{2} \quad \quad j=\{1,2,3,\cdots, N\}.
\end{equation}
Such that the above equation Eq.\eqref{combinedeqn2} becomes
\begin{equation}
    \frac{\lambda_j+d+\frac{i}{2}}{\lambda_j+d-\frac{i}{2}} \frac{\lambda_j-d+\frac{i}{2}}{\lambda_j-d-\frac{i}{2}}\left(\frac{\lambda_\alpha-b+i}{\lambda_\alpha-b-i} \right)^N\left(\frac{\lambda_\alpha+b+i}{\lambda_\alpha+b-i} \right)^N=\prod_{\beta\neq\alpha} \frac{\lambda_\alpha-\lambda_\beta+i}{\lambda_\alpha-\lambda_\beta-i}\frac{\lambda_\alpha+\lambda_\beta+i}{\lambda_\alpha+\lambda_\beta-i},
\end{equation}
and the equation for quasi-momentum Eq.\eqref{quasimomrell} becomes
\begin{equation}
e^{-2ik_j L}=\prod_{\alpha=1}^M \left(\frac{b+\lambda_\alpha+i}{b+\lambda_\alpha-i} \right) \left(\frac{b-\lambda_\alpha+i}{b-\lambda_\alpha-i} \right).
\end{equation}

These are exactly the same as the Bethe equations we obtained above by using the fusion as the level of $R-$matrices and transfer matrices rather than at the level of Bethe equations.

As explained in the main text, the total number of spin-$\frac{1}{2}$ particles is always even in the bulk as it is constructed from two copies of $SU(2)_1$ Gross-Neveu model with equal numbers of particles. When a spin-$\frac12$ impurity is added to the boundary, the total number of spin-$\frac{1}{2}$ particles is always odd. Hence, there is a propagating spinon of spin-$\frac12$ that is present in the ground state of the model. Moreover, it is important to notice that since the model is integrable, there is no pair production during the scattering process, and hence, it makes sense to talk about the model defined on an open line with a fixed number of particles.

\section{Thermodynamics in the Kondo phase}\label{sec:TBA}

Under string hypothesis one assumes that all solutions of  Eq.\eqref{BAEspin1/2Wspin1/2} are string solutions of the form 
\begin{equation}
    \lambda_j^{(n)} = \lambda^{(n)} + i\frac{1}{2}(n + 1 - 2j), \quad j = 1, 2, \ldots, n.
\end{equation}

Plugging in these solution in the Bethe equations Eq.\eqref{BAEspin1/2Wspin1/2} upon taking $\ln$ on both sides become
\begin{equation}
  \Theta_{n}(\lambda^{(n)}_\gamma)+\sum_\nu  \Theta_{n}(\lambda^{(n)}_\gamma+\nu  d)+N \Theta_{n-1}(\lambda^{(n)}_\gamma+\nu  b)+N\Theta_{n+1}(\lambda^{(n)}_\gamma+\nu  b)=\sum_{m,\beta}\Theta_{n,m}(\lambda^{(n)}_\gamma+\nu \lambda^{(m)}_\beta)-2\pi I_\gamma^{(n)},
\end{equation}

where
\begin{equation}
    \Theta_n(x)=-2 \tan ^{-1}\left(\frac{2 x}{n}\right),
\end{equation}

and

\begin{equation}
    \Theta_{mn}(x) = \begin{cases} 
\Theta_{|n-m|}(x) + 2\Theta_{|n-m|+2}(x) + \cdots + 2\Theta_{n+m-2}(x) + \Theta_{n+m}(x), & n \neq m \\ 
2\Theta_2(x) + \cdots + 2\Theta_{2n-2}(x) + \Theta_{2n}(x), & n = m .
\end{cases}
\end{equation}

The counting function 
\begin{equation}
    \begin{aligned}
        \nu_n(\lambda) = \frac{1}{2\pi}\left[\Theta_{n}(\lambda^{(n)}_\gamma)+\sum_\nu  \Theta_{n}(\lambda^{(n)}_\gamma+\nu  d)+N \Theta_{n-1}(\lambda^{(n)}_\gamma+\nu  b)+N\Theta_{n+1}(\lambda^{(n)}_\gamma+\nu  b)-\sum_{m,\beta}\Theta_{n,m}(\lambda^{(n)}_\gamma+\nu \lambda^{(m)}_\beta) \right]
    \end{aligned}
\end{equation}
is such that it gives the integers $I^{(n)}_\gamma$ for corresponding roots $\lambda^{(n)}_\gamma$ \textit{i.e.} $\nu_n(\lambda^{(n)}_\gamma)=I^{(n)}_\gamma$ and gives skipped integers $I^{(n),h}_\gamma$ are the positions of holes \textit{i.e.} $\nu_n(\lambda^{(n),h}_\gamma)=I^{(n),h}_\gamma$.          
The derivative of the counting function in the thermodynamic limit gives the density of $n$-strings $\sigma_n(\mu)$ and holes $\sigma_n^h(\mu)$
\begin{equation}
    \frac{\mathrm d \nu_n}{\mathrm d \mu} = \sigma_n(\mu) + \sigma_n^h(\mu).
\end{equation}

Combining the last two expressions gives
\begin{equation}\label{sigma_n^h}
    \sigma_n^h(\mu) = f_n(\mu) - \sum_{m=1}^{\infty}A_{nm} \sigma_m(\mu).
\end{equation}
The notations used above are
\begin{equation}
    \begin{aligned}
        f_n(\mu)&=N K_{n+1}(\mu- b)+N K_{n-1}(\mu- b)+N K_{n+1}(\mu+ b)+N K_{n-1}(\mu+ b) +K_{n}(\mu+ d) +K_{n}(\mu- d)+K_{n}(\mu)\\
        A_{nm}&=\left[|n-m|\right]+2\left[|n-m|+2\right]+\ldots+2\left[n+m-2\right]+\left[n+m\right],
    \end{aligned}
\end{equation}
with $K_n(\mu)$ defined as
\begin{equation}
    K_n(\mu) \equiv -\frac{1}{2\pi} \frac{\mathrm{d}\Theta_n}{\mathrm{d}\mu} 
    = \frac{1}{\pi} \frac{\frac{n}{2}}{\left(\frac{n}{2}\right)^2 + \mu^2},
\end{equation}

and functional $\left[ n\right]$ introduced as convolution with $K_n$:
\begin{equation}
    \left[n\right] g(\mu)\equiv K_n \star g(\mu) = \int \mathrm{d} \lambda \,K_n(\mu-\lambda)g(\lambda).
\end{equation}

In terms of the string variables $\lambda^{(n)}$, the energy function (obtained by summing all quasimomenta $\sum_j k_j$ in Eq.\eqref{CBAEapp}) can be expressed as
\begin{align}
    E=&\sum_j\frac{2\pi}{L}n_j+\sum_n D\int\mathrm{d}\lambda \sigma_n(\lambda)\left[ \Theta_{n-1} \left(  b-\lambda\right)+\Theta_{n+1} \left(  b-\lambda\right)+\Theta_{n-1} \left(\lambda+  b\right)+\Theta_{n+1} \left(\lambda+  b\right)-4\pi\right].
\end{align}

Here, the first term is the charge energy
\[
E^{(c)}(\{n_j\}) = \frac{2\pi}{L} \sum_{j=1}^{N^e} n_j,
\]
such that the charge partition function becomes
\[
Z^{(c)} = \sum_{\{n_j\},\, n_j \geq -N^e} \exp\left[-\frac{1}{T} \sum_{j=1}^{N^e} \frac{2\pi}{L} n_j \right].
\]
This describes the thermodynamics of \(N\) noninteracting spinless fermions with linear kinetic energy. In the limit \(D \to \infty\), it leads to the free energy
\[
F^{(c)} = - \frac{L T}{2 \pi} \int_{-\infty}^\infty dk \ln \left(1 + e^{-\frac{k}{T}} \right) = -\frac{\pi}{12} L T^2 + \{\text{infinite constant}\}.
\]
This free energy corresponds to half the free energy of a noninteracting electron gas at zero magnetic field.

The free energy of the spin part in the presence of the magnetic field $H$ can be written as
\begin{equation}
    \mathcal{F}=E+2M H - T\mathcal{S},
\end{equation}
where $\mathcal{S}$ is the Yang-Yang entropy, which, upon using the Stirling approximation, can be written as
\begin{equation}
    \mathcal{S}=\sum_{n=1}^\infty \int \mathrm{d} \lambda\, \left[ (\sigma_n+\sigma_n^h)\ln(\sigma_n+\sigma_n^h)-\sigma_n\ln\sigma_n - \sigma_n^h\ln\sigma_n^h \right].
\end{equation}
Combining $E+M H$ as
\begin{equation}
    E+M H =\sum_{n=1}^\infty \int \mathrm{d} \lambda \, g_n(\lambda) \sigma_n(\lambda),
\end{equation} 
and introducing $$g_n(\lambda)=2nH+D\left[ \Theta_{n-1} \left(  b-\lambda\right)+\Theta_{n+1} \left(  b-\lambda\right)+\Theta_{n-1} \left(\lambda+  b\right)+\Theta_{n+1} \left(\lambda+  b\right)-4\pi \right],$$ one can write the free energy as
\begin{equation}\label{F}
    \mathcal{F}=\sum_{n=1}^\infty \int \mathrm{d} \lambda \, \left[ g_n \sigma_n - T\sigma_n \ln\left[1+\frac{\sigma_n^h}{\sigma_n}\right] - T\sigma_n^h \ln\left[1+\frac{\sigma_n}{\sigma_n^h}\right] \right].
\end{equation}
Varying the free energy subjected to the constrain $\delta \sigma_n^h = -\sum_{m=1}^\infty A_{nm}\delta \sigma_m$  from Eq.\eqref{sigma_n^h} we get
\begin{equation}
    g_n-T\ln\left[1+\frac{\sigma_n^h}{\sigma_n}\right]+T\sum_{m=1}^\infty A_{nm}\ln\left[1+\frac{\sigma_m}{\sigma_m^h}\right]=0.
\end{equation}
or, introducing $\eta_n = \sigma_n^h/\sigma_n$, one can write
\begin{equation}\label{pre TBA}
\ln\left[1+\eta_n(\lambda)\right]=\frac{g_n(\lambda)}{T}+\sum_{m=1}^\infty A_{nm} \ln\left[1+\eta^{-1}_m(\lambda)\right]\,.
\end{equation}

It is convenient to introduce a functional $G$ acting by convolution with $1/2\cosh(\pi \lambda)$
\begin{equation}
    Gf(\lambda) = \int \mathrm{d} \mu \, \frac{1}{2\cosh\pi (\lambda-\mu)}f(\mu).
\end{equation}

Applying $\delta_{m,n}-G(\delta_{m-1,n}+\delta_{m+1,n})$ to the Eq.\eqref{pre TBA}, we obtain
\begin{equation}\label{TBA}
    \ln \eta_n(\lambda) = 
   -\frac{m}{T}\cosh(\pi \lambda)\delta_{n,2}+G\ln\left[ 1+\eta_{n+1} \right]+G\ln\left[ 1+\eta_{n-1} \right].
\end{equation}

Applying $G$ on Eq.\eqref{pre TBA}, we obtain
\begin{align}
    G\left[\ln[1+\eta_n(\lambda)]-\frac{g_n(\lambda)}{T} \right]=\sum_{m=1}^\infty Y_{n,m}\ln[1+\eta_n^{-1}(\lambda)]
\end{align}
where we introduced 
    \begin{equation}
        Y_{n,m}(\mu)=\sum_{l=1}^{\mathrm{min}(n,m)} K_{n+m+1-2l}(\mu).
\end{equation}
Noticing
\begin{equation}
    Y_{2,m}(\mu)=K_{n-1}(\mu)+K_{n+1}(\mu),
\end{equation}
and
\begin{equation}
    Y_{1,m}(\mu)=K_m(\mu),
\end{equation}
we can simplify the equation for free energy as 
\begin{equation}
    \begin{aligned}
        \mathcal{F}&=\frac{1}{2}\int \mathrm{d} \lambda\, \left\{ \left( \frac{N}{2\cosh \pi (\lambda-b)}+\frac{N}{2\cosh \pi (\lambda+b)}\right)\left[ g_{2}(\lambda)-T\ln(1+\eta_{2}(\lambda)) \right]\right.\\
        &\left.+\left( \frac{1}{2\cosh\pi \lambda}+\frac{1}{2\cosh\pi (\lambda-d)}+\frac{1}{2\cosh\pi (\lambda+d)} \right)\left[g_1(\lambda)-T\ln(1+\eta_1(\lambda))\right] \right\}.
    \end{aligned}
\end{equation}
The impurity part of the free energy is
\begin{equation}
    \begin{aligned}
        \mathcal{F}_{\mathrm{imp}}&=\mathcal{F}_{\mathrm{imp}}^0-\frac{T}{2}\int \mathrm{d} \lambda\,\left(\frac{1}{2\cosh\pi (\lambda-d)}+\frac{1}{2\cosh\pi (\lambda+d)} \right)\ln(1+\eta_1(\lambda)).
    \end{aligned}
    \label{freeenegg}
\end{equation}
These are the equations studied in the main text.

\section{Derivation of renormalization group equations}\label{RGderivation}
In the main text, we obtained the expression for the superconducting mass gap
\begin{equation}
m=D \arctan\left(\frac{1}{\sinh(\pi b)}\right).
\end{equation}
Upon taking the scaling limit $D\to \infty$ and $b\to \infty$ while holding $m$ fixed, we write
\begin{equation}
    m=2De^{-\pi b}= 2De^{-\frac{\pi}{2g}}.
\end{equation}
Inverting this relation, we obtain
\begin{equation}
    \frac{1}{2b(D)}=g(D)=\frac{\pi }{2 \ln \left(\frac{2 D}{m}\right)}.
    \label{gflow}
\end{equation}
Moreover, from the expression of the RG invariant quantity $d$, we obtain
\begin{equation}
    c=\frac{8 b}{4 b^2-4 d^2-9},
\end{equation}
which gives
\begin{equation}
    J(D)=\frac{18-8 b^2+8 d^2\pm \sqrt{\left(18-8 b^2+8 d^2\right)^2+512 b^2}}{32 b} .
    \label{jflow}
\end{equation}

Using Eq.~\eqref{gflow} in Eq.~\eqref{jflow} and differentiating Eq.~\eqref{gflow} and Eq.~\eqref{jflow} with respect to $\ln D$, we arrive at the RG equations
\begin{align}
    \beta(g)=\frac{\mathrm{d}}{\mathrm{d}\ln D}g&=-\frac{2}{\pi}g^2,\\
    \beta(J)=\frac{\mathrm{d}}{\mathrm{d}\ln D}J&=-\frac{2 J \left(2 g J^2-g+J\right)}{\pi  \left(2 J^2+1\right)}.
\end{align}
\end{document}